\documentclass[12pt,reqno]{amsart}

\usepackage[samelinetheorem,notitle]{maherart}

\hypersetup{
    pdftitle =
        {Matching with Generalized Sequential Reserves},
    pdfauthor =
        {Orhan Aygün and Bertan Turhan
        },
    pdfsubject=
        {Sequential Reserves}
}

\usepackage{tabularx}
\usepackage{multirow}
\usepackage{longtable}
    \newcolumntype{L}{>{\raggedright\arraybackslash}X}

\usepackage{xcolor}

\usepackage{nicematrix,booktabs,caption}

\begin{document}
\raggedbottom
\author[Aygün]{Orhan Aygün$^\textrm{1}$}
\address{$^\textrm{1}$ Boğaziçi University}

\author[Turhan]{Bertan Turhan$^\textrm{2}$}
\address{$^\textrm{2}$ Iowa State University}

\date{\today}

\title{Matching with Generalized Sequential Choice Rules}
\thanks{This paper subsumes and replaces the earlier working paper titled \textit{``Matching with Generalized Lexicographic Choice Rules}.'' We are grateful to the editor, Faruk Gul, the associate editor, and the two reviewers for their constructive and helpful feedback. We also thank Itai Ashlagi, Eduardo Azevedo, David Delacrétaz, Battal Do\u{g}an, Federico Echenique, Aytek Erdil, Isa Hafal\i r, Onur Kesten, Aditya Kuvalekar, Kriti Manocha, Debasis Mishra, Josué Ortega, Assaf Romm, Al Roth, Arunava Sen, Bumin Yenmez, Kemal Yildiz, as well as the participants at the ISU Market Design Conference, the 2022 Asian Meeting of the Econometric Society in East and Southeast Asia, the 2022 Conference on Mechanism and Institution Design, the 2024 North American Meeting of the Econometric Society, and the 2024 Asian Meeting of the Econometric Society for their helpful comments and suggestions.}

\begin{abstract}
\onehalfspacing
This paper studies a many-to-one matching between individuals and institutions where institutions comprise multiple divisions and face cross-divisional constraints. We introduce a parametrized family of choice rules, which we call \textit{generalized sequential} (GSq), that encompasses many different choice rules encountered in practice and in market design literature. We show that the cumulative offer mechanism (COM) is the unique stable and strategy-proof mechanism in a matching problem where all institutions have a GSq choice rule. We present two real-world applications in which choice rules in the GSq family emerge naturally: affirmative action in India's public higher educational institutions and government jobs and high school admissions in China. 

\end{abstract}

\begin{titlepage}
    \maketitle
\end{titlepage}

\section{Introduction}

Many real-life allocation problems involve matching individuals with institutions that comprise multiple divisions (or reserve categories), such as firms, hospitals, or schools. In rationing healthcare resources (\cite{aziz2024efficient}, \cite{pathak2024fair}, and \cite{evren2025reserve}) and the allocation of public goods such as seats in prestigious engineering schools (\cite{baswanaetal2019}, \cite{aygun2022dereserve}) and government jobs (\cite{sonmez2022affirmative}) available resources are divided into different reserve categories, each with a subset of individuals eligible for that category.  Divisions are endowed with their own choice rules. 

In some applications, divisions' choice rules are induced by a strict ranking of individuals and always select the q-best individuals, whenever available, where $q$ denotes the capacity. Such choice rules are called \textit{q-responsive}. Similarly, in many rationing practices, resource units are assigned according to a baseline priority order and the eligibility of candidates in each reserve category. 

However, in many other allocation problems, divisions' choices are more complicated. To be concrete, consider a division with two positions that selects applicants based on merit scores and requires at least one position to be filled with a woman applicant. Suppose that the two top-scoring applicants are male, followed by a female applicant. This division selects the male applicant with the highest score, together with the female applicant, to meet the gender requirement. This kind of choice rule is not responsive because the two highest-scoring applicants were not chosen.

Often, divisions are interconnected in that the number of positions available in one division depends on how many applicants other divisions hire. This creates cross-divisional constraints across the institution. For example, consider a high school that reserves slots for two middle schools in the same catchment area, while the remaining slots are open to all candidates, including students from other areas. Suppose that the school first fills the reserved slots for the two middle schools, followed by open slots. If there are not enough applicants from any of the middle schools, the remaining positions reserved for the middle school are allocated as open positions.\footnote{This is an example of a \textit{soft} reserve policy. There is also a \textit{hard} reserve policy in which the remaining reserved positions are not made available to others; See \cite{hafalir2013effective} and \cite{ehlers2014school} for a detailed discussion on hard and sort reservation policies.} As such, the capacity of the "open" positions depends on the number of vacancies in the reserved slots for the two middle schools.  \cite{roth1999redesign} were the first to address interdivisional constraints of this kind in matching theory. They introduced the notion of ``reversion" in the context of the National Residency Matching Program (NRMP), which matches doctors with hospital programs. Reversion is a practice in which unfilled positions in one hospital program can be transferred to other programs, while respecting the hospital's overall capacity limits. In their analysis, \cite{roth1999redesign} identified reversion as a significant complication in the NRMP's matching process.

A significant body of market design literature examines the design and implementation of soft reserve policies in rationing and resource allocation problems. The specific method of soft reservation implementation has important distributional consequences. In the context of admissions to technical universities in India, \cite{aygun2022dereserve} introduced a soft reserve policy that reverts unfilled slots reserved for particular groups to the open category, which is filled before reserve categories by updating the position categories' capacities and re-running the choice rule from scratch. In a similar framework, \cite{aygun2025affirmative} study forward transfer choice rules in which reserved slots that are not filled are allocated as an open category after the reserve categories are filled. \cite{hu2025verifiable} presents various soft reserve policies currently in use for high school admissions in China. \cite{pathak2024fair} introduced a smart reserve matching algorithm in the context of allocating life-saving medical resources. 

This paper examines the choice problem faced by multi-divisional institutions that encounter cross-divisional constraints. We define a family of choice rules, which we call generalized sequential (GSq), that encompasses many different choice rules encountered in practice and in market design literature. We show that the generalized deferred acceptance (DA) is the unique stable and strategy-proof mechanism in a matching problem where all institutions have a GSq choice rule.\footnote{In our setting, the term "sequential" in GSq choice rules does not refer to a choice model with time component, as there is no time component in our framework. We consider static choice rules.}

The GSq choice rules allow divisions to use substitutable and size-monotonic choice rules rather than constrain them to simpler $q$-responsive choice rules. In addition, these rules facilitate a more comprehensive soft reserve policies, as defined through \textit{transfer functions}. Specifically, we present a parameterized family of choice rules constructed from substitutable choice rules that satisfy size monotonicity — two of the most prominent properties in the market design literature. 

The choice rules in the GSq family have the following structure: There is a predetermined \textbf{order of precedence} over divisions according to which the institution fills the positions of its divisions. Each division has an initial number of positions available that may be updated and is endowed with a series of choice rules, one for each (possible) integer capacity. The choice of a division depends on the set of available options (or contracts) and its capacity, both of which are determined by the choices made by the preceding divisions. The set of available options (or contracts) can be thought of as the set of remaining options from the choices of preceding divisions. The capacity of each division is a function of the number of unfilled positions in the preceding divisions and is given by an exogenously specified \textbf{transfer function}. The \textbf{aggregate choice rule} of an institution is then defined as the union of its divisions' choice rules.  

The choice rule for each division satisfies the \textit{substitutes} and the \textit{size monotonicity} properties for each possible integer capacity. The substitutes property (\cite{kelso1982job} and \cite{gul1999walrasian}) and size monotonicity (\cite{alkan2003stable} and \cite{hatfield2005matching}) impose restrictions on the choice function when the set of available alternatives changes for a given capacity. A choice rule satisfies the substitute property if there are no two alternatives $x$ and $y$ that are complementary, in the sense that gaining access to $x$ makes $y$ desirable. A choice rule is size monotonic if it chooses weakly more alternatives whenever the set of available alternatives expands. 

In our framework, a division's capacity may be increased if there are vacancies in the preceding divisions. We, therefore, need to specify how a division can select from any given set of options under different capacities. When the capacity of a division increases, the division's choice rules with initial and increased capacities can be regarded as two different choice rules. These choice rules are closely connected via the \textit{quota monotonicity} property, which is similar to the notion of \textit{expansion} defined in \cite{chambers2017choice}.\footnote{The quota monotonicity property requires consistency in choices when the capacity increases. It is a stronger condition than the expansion property of \cite{chambers2017choice}.} The choice rule with the increased capacity is an \textit{expansion} of the choice rule with the initial capacity. The quota monotonicity property is satisfied if 
\begin{itemize}
	\item all alternatives selected under the initial capacity are chosen when the capacity increases, and 
	\item the difference between the number of chosen alternatives under the increased and initial capacities cannot exceed the increase in capacity. 
\end{itemize}

We formulate cross-division constraints via transfer functions. Given the order of precedence over the divisions and capacity of the first division, the capacity of a division is a function of the number of vacancies in the divisions that precede it. We require that the transfer function satisfies a mild condition, called monotonicity (\cite{westkamp2013analysis}). A transfer function is \textbf{monotone} if:
\begin{enumerate}
    \item whenever there are weakly more seats unfilled in every division preceding the $j^{th}$ division, weakly more slots should be available for the $j^{th}$ division, and 
    \item the capacity increase at each division cannot exceed the total number of vacant slots from the divisions preceding it. That is, no additional seats are created.
\end{enumerate}
  Condition (1) requires that the capacity function of each division is weakly increasing. Note that condition (2) is aggregate across divisions up to $j$, rather than a direct bound on the capacity change of a single division. 

An aggregate choice rule is in the GSq family if divisions' choice rules satisfy the substitutes property, size monotonicity, and quota monotonicity, and the transfer function is monotonic. Aggregate choice rules in the GSq family may fail to satisfy the substitutes property and size monotonicity. In Theorem \ref{Theorem 1}, we show that each aggregate choice rule in the GSq family satisfies a novel condition defined by \cite{hatfield2019hidden}. An immediate corollary of Theorem \ref{Theorem 1} is obtained when combined with the results from \cite{hatfield2017restud} and \cite{hatfield2019hidden}: The generalized deferred acceptance (DA) mechanism is the unique stable mechanism that is strategy-proof.

In practice, policymakers frequently implement choice rules characterized by intuitive definitions. The general classes of choice rules can be too complex to articulate effectively (\cite{echenique2007counting}). Choice rules in the GSq family appear naturally in important real-world applications. For example, policy goals and affirmative action restrictions in India uniquely identify a particular choice function for reserve categories, the \textbf{two-step meritorious horizontal choice rules} (\cite{sonmez2022affirmative}). We show that an aggregate choice rule of an institution, where each division employs the two-step meritorious horizontal choice rule, is in the GSq family under any monotone transfer function (Proposition \ref{Prop1}). In addition, various choice rules that are being implemented and a new one proposed by \cite{hu2025verifiable} for admission to high schools in China are also in the GSq family (Propositions 2-6).

The GSq family nests other prominent choice rules introduced in the market design literature, including the reserve-based choice rules of \cite{echenique2015control}, slot-specific priority choice rules of \cite{kominers2016matching}, dynamic reserves choice rules of \cite{aygun2020dynamic} and multi-price choice rule of \cite{greenberg2024redesigning}, among many others. GSq choice rules offer considerable latitude to integrate complex diversity constraints. Moreover, the general framework we introduce is flexible to accommodate many other real-world matching applications, such as airline seat upgrades (\cite{kominers2016matching}), multidimensional reserves in Brazil (\cite{aygun2021college}), high school admissions in China (\cite{hu2025verifiable}), affirmative action implementations in Chilean school choice (Correa et al. (2021)), among others. Thus, the GSq family provides a unified framework for market designers.

\subsection{Literature Review} \label{sec:literature}

\cite{hatfield2017stable} also study matching markets, where institutions have multiple divisions that face capacity constraints similar to ours. The authors also study how hiring in one division affects the resources available for hiring in other divisions. Unlike our GSq choice rules, the complexity of choice rules with flexible allotments they introduce could make them challenging to implement in practice. Their allotment functions require determining how many positions are allocated to each division given the available set of alternatives. 

This paper contributes to the existing literature on diversity constraints in matching market design. The idea of reserve was introduced in the seminal work by \cite{hafalir2013effective} after \cite{kojima2012school} pointed out the drawbacks of the quota policy. \cite{ehlers2014school} show that, with hard bounds, there might not exist assignments that satisfy standard fairness and non-wastefulness properties.  To achieve fair and non-wasteful assignments, the authors propose to interpret constraints as soft bounds. There is now a rich market design literature on implementing diversity in matching markets. Some other examples are \cite{kamada2015efficient}, \cite{biro2020need}, and \cite{hassidim2021limits}.

This paper contributes to the growing body of literature on resource allocation problems in India. In Section 5.1, we examine resource allocation problems in India under complex affirmative action constraints as an application. We show that employing the meritorious horizontal choice rule (\cite{sonmez2022affirmative}) within each division in conjunction with monotone transfer functions produces choice rules within the GSq family. Our formulation explicitly accounts for the de-reservation of vacant Other Backward Classes (OBC) positions, a consideration absent from \cite{sonmez2022affirmative}.\footnote{\cite{aygun2022dereserve} examine the concurrent implementation of vertical reservations and OBC de-reservations, specifically focusing on the recently revamped admissions procedures for technical universities (IITs) in India, as originally designed and implemented by \cite{baswanaetal2019}. See also \cite{aygun2023priority} for computationally simpler than and outcome equivalent to backward transfers choice rule.} Furthermore, our model allows individuals to express preferences over pairs of institution-position category, improving the applicability of the framework to real-world allocation scenarios. \cite{echenique2015control} is the first paper to present Indian affirmative action as an example of controlled school choice. Subsequent work by \cite{aygun2017large} delves into the specific challenges in admissions to IITs. 

\cite{aygun2020dynamic} introduced a family of \textit{dynamic reserves} choice rules, which were shown to nest the slot-specific priorities choice rules of \cite{kominers2016matching}. The family of GSq choice rules nests dynamic reserve choice rules. This generalization is significant, as dynamic reserve choice rules were formulated using q-responsive divisional choice rules, which are inherently limited in incorporating horizontal reservations considered in \cite{sonmez2022affirmative}. 

The rest of the paper is organized as follows. Section \ref{Model} formulates the model of many-to-one matching with contracts. Section \ref{GSq} introduces the family of GSq choice rules that are central to our analysis.  Section \ref{MecDesign} presents the main result of the paper on stable and strategy-proof mechanisms through GSq choice rules. Section \ref{Applications} presents two real-world applications in which GSq choice rules are used or proposed: affirmative action in India and high school admissions in China. Finally, Section \ref{sec:concl} concludes. All formal proofs are presented in the appendix.

\section{Preliminaries} \label{Model}

We consider a centralized many-to-one matching market. There is a finite set of \textit{individuals} $I=\{i_{1},...,i_{n}\}$ and a finite set of \textit{institutions} $S=\{s_{1},...,s_{m}\}$. We let $\overline{q}_{s}$ be the capacity of the institution $s$. 

There is a finite set of contracts $X$. Each contract $x\in X$ is between an individual $\mathbf{i}(x)$ and an institution $\mathbf{s}(x)$. For any set of contracts $Y\subseteq X$, let $\mathbf{i}(Y)\equiv \bigcup_{x\in Y}\{\mathbf{i}(x)\}$ be the set of individuals who have a contract associated with them in $Y$. Similarly, we let $\mathbf{s}(Y)\equiv \bigcup_{x\in Y}\{\mathbf{s}(x)\}$. We also define $Y_{i}\equiv \{x\in Y\mid i=\mathbf{i}(x)\}$  for each $i\in I$ and $Y_{s}\equiv \{x\in Y\mid s=\mathbf{s}(x)\}$ for each $s\in S$. 

Each individual $i\in I$ has a strict preference $P_{i}$ over contracts
in $X_{i}=\{x\in X | \mathbf{i}(x)=i\}$ and an outside option $\emptyset$. We denote $\mathcal{P}_{i}$ the set of all strict preferences individual $i$ may have. A contract $x\in X_{i}$ is acceptable to $i$ with respect to $P_{i}$ if $xP_{i}\emptyset$. We let $R_{i}$ be the weak preference relation associated with $P_{i}$. $xR_{i}y$ if and only if $xP_{i}y$ or $x=y$.

Each institution $s\in S$ has \textit{multiunit} demand and is endowed with an \textbf{aggregate choice rule} $C_{s}$ that describes how $s$ would choose from any set of contracts. For all $Y\subseteq\ X$ and $s\in S$, the aggregate choice rule $C_{s}$ is such that $C_{s}(Y)\subseteq Y_{s}$ and $\mid C_{s}(Y)\mid \leq \overline{q}_{s}$. In the next section, we introduce a rich class of choice rules for institutions.

A matching is a set of contracts $Y\subseteq X$. A matching $Y\subseteq X$ is \textbf{feasible} if $\mid Y_{i}\mid\leq1$ for all $i\in I$. A feasible matching $Y\subseteq X$ is \textbf{stable} if 
\begin{enumerate}
    \item $Y_{i}R_{i}\emptyset$, for all $i\in I$, 
	\item $C_{s}(Y)=Y_{s}$, for all $s\in S$, and
	\item there is no $Z\subseteq(X \setminus Y)$, such that $Z_{s}\subseteq C_{s}(Y\cup Z)$ for all $s\in\mathbf{s}(Z)$ and $Z_{i}P_{i}Y_{i}$ for all $i\in\mathbf{i}(Z)$. 
\end{enumerate}

 Condition (1) is the standard individual rationality condition for individuals. Condition (2) requires that institutions' selection procedures be respected and can be interpreted as an individual rationality condition for institutions.\footnote{This condition ensures that criteria, such as diversity and affirmative action, that are encoded in institutions' choice rules are complied with; See \cite{alva2023choice} for a detailed discussion of this property.} Condition (3) is the standard no blocking condition and requires that there be no set of contracts $Z$ such that all institutions and individuals associated with contracts in $Z$ prefer to receive the contracts in $Z$.  

Stability and strategy-proofness are fundamental objectives in market design. Stability precludes scenarios in which agents can profitably rearrange their matches after the market has cleared (\cite{gale1962college}). Stability is considered a crucial determinant of the success of a centralized clearinghouse (\cite{roth1990new}). Stability also ensures that the institutional objectives, typically set by policymakers and encoded in institutions' choice rules, are satisfied. Moreover, there is a sense in which stability implies fairness (or no justified envy).\footnote{See \cite{romm2024stability} for a detailed analysis of the relationship between stability and no justified envy in a matching model with contracts.} Strategy-proofness establishes truthful preference revelation as a dominant strategy, eliminating incentives for strategic misrepresentation, and thus simplifying market participation while ensuring that allocations derive from accurate preference data.

Given a profile of aggregate choice rules of institutions $C=(C_{s})_{s\in\ S}$, a mechanism $\psi(\cdot;C)$ maps preference profiles $P=(P_{i})_{i\in\ I}$ to matchings. Unless otherwise stated, we assume that institutions' choice rules are fixed. A mechanism $\psi$ is \textbf{stable} if the outcome of $\psi$, $\psi(P;C)$, is a stable matching for every preference profile $P$, given a profile of the institutions' choice rules $C$. A mechanism $\psi$ is \textbf{strategy-proof} if, given a profile of institutions' choice rules $C$, for every preference profile $P$ and for each individual $i\in I$,
there is no $\widetilde{P}_{i}\in \mathcal{P}_{i}$ such that 

$$\psi(\widetilde{P}_{i},P_{-i};C)P_{i} \psi(P;C).$$ 

\section{Generalized Sequential Choice Rules} \label{GSq}

This section introduces a class of choice functions, named \textit{Generalized Sequential} (GSq). Consider an institution $s$ that has a set of \textit{divisions} $\mathcal{K}_{s}=\{1,...,K_{s}\}$.
Each division $k\in\mathcal{K}_{s}$ has an associated \textit{choice} rule $C_{s}^{k}: 2^{X} \times\mathbb{Z}_{+}\longrightarrow2^{X_{s}}$
that specifies the contracts that division $k$ chooses given a set of contracts and a \textit{capacity}. The capacity of division $k>1$ may vary as a function of the number of vacant positions in the preceding divisions. We require that for a set of contracts $Y\subseteq\ X_{s}$ and a capacity of $\lambda \in \mathbb{Z}_{+}$, $\mid C_{s}^{k}(Y;\lambda)\mid\leq\lambda$ and that

$$
x,x^{'}\in C^{k}_{s}(Y;\lambda) \implies \mathbf{i}(x) \neq \mathbf{i}(x^{'}).
$$
That is, the choice rule of a division can select \textit{at most one} contract for an individual.

Given a set of contracts $Y\equiv Y^{1}\subseteq\ X$ , a capacity $\overline{q}_{s}$ for institution $s$, and a capacity $\overline{q}_{s}^{1}$ for division 1, we compute the set chosen $C_{s}(Y,\overline{q}_{s})$ sequentially in $K_{s}$ steps, where division $k$ chooses in step $k$, for $k=1,...,K_{s}$,
as follows: 

\paragraph{Step 1}
Given $Y^{1}$ and $\overline{q}_{s}^{1}$, division 1 chooses $C_{s}^{1}(Y^{1};\overline{q}_{s}^{1})$. Let
\[
r_{1}=\overline{q}_{s}^{1}-\mid C_{s}^{1}(Y^{1};\overline{q}_{s}^{1})\mid
\]
be the number of vacant positions in division 1. Moreover, let
\[
Y^{2}\equiv Y^{1}\setminus\{x\in Y^{1}\mid\mathbf{i}(x)\in\mathbf{i}[C_{s}^{1}(Y^{1};\overline{q}_{s}^{1})]\}
\]
be the set of contracts of individuals who were not selected by division 1. 

\paragraph{Step k ($2\protect\leq k\protect\leq K_{s})$}
The capacity of division $k$ is considered as a function of the number of unfilled positions in the preceding divisions, $q_{s}^{k}(r_{1},...,r_{k-1})$, since we allow redistribution of vacant positions in the preceding divisions.\footnote{We give the details of capacity redistributions in the next subsection, titled transfer policy.}

Given the set of remaining contracts $Y^{k}$ and division $k$'s capacity $q_{s}^{k}(r_{1},...,r_{k-1})$, division $k$ chooses
$C_{s}^{k}(Y^{k};q_{s}^{k}(r_{1},...,r_{k-1}))$. Let 
\[
r_{k}=q_{s}^{k}(r_{1},...,r_{k-1})-\mid C_{s}^{k}(Y^{k};q_{s}^{k}(r_{1},...,r_{k-1}))\mid
\]
be the number of unfilled positions in division $k$. Moreover, let 
\[
Y^{k+1}\equiv Y^{k}\setminus\{x\in Y^{k}\mid\mathbf{i}(x)\in\mathbf{i}[C_{s}^{k}(Y^{k};q_{s}^{k}(r_{1},...,r_{k-1}))]\}
\]

be the set of contracts of individuals who were not selected by any
division in the first $k$ steps. 

The union of divisions' choices is "the" institution's chosen set.
That is, 
\[
C_{s}(Y,q_{s})\equiv C_{s}^{1}(Y^{1};\overline{q}_{s}^{1})\cup \bigcup_{k=2}^{K_{s}} C_{s}^{k}(Y^{k};q_{s}^{k}(r_{1},...,r_{k-1})).
\]

\subsection*{Transfer Policy}

Given an initial capacity of the first division $\overline{q}_{s}^{1}$,
a \textit{transfer policy} of institution $s$ is a sequence of capacity functions $q_{s}=(\overline{q}_{s}^{1},(q_{s}^{k})_{k=2}^{K_{s}})$,
where $q_{s}^{k}:\:\mathbb{Z}_{+}^{k-1}\longrightarrow\mathbb{Z}_{+}$
for all $k\in\mathcal{K}_{s}$ and such that 
\[
\overline{q}_{s}^{1}+q_{s}^{2}(0)+q_{s}^{3}(0,0)+\cdots+q_{s}^{K}(0,...,0)=\overline{q}_{s}.
\]

We impose the following mild condition on transfer functions.

\begin{definition}
    A transfer policy $q_{s}$ is $\mathbf{monotonic}$
if, for all $j\in\{2,...,K_{s}\}$ and all pairs of sequences $(r_{l},\widetilde{r}_{l})$,
such that $\widetilde{r}_{l}\geq r_{l}$ for all $l\leq j-1$, 
\begin{equation}
    q_{s}^{j}(\widetilde{r}_{1},...,\widetilde{r}_{j-1})\geq q_{s}^{j}(r_{1},...,r_{j-1});
\end{equation}
and
\begin{equation}
    \sum_{m=2}^{j} [q_{s}^{m}(\tilde{r}_{1},...,\widetilde{r}_{m-1})-q_{s}^{m}(r_{1},...,r_{m-1})]\leq\sum_{m=1}^{j-1}[\widetilde{r}_{m}-r_{m}].
\end{equation}
\end{definition}

The first condition simply says that the capacity of a division is a (weakly) increasing function of the number of vacancies in the preceding divisions. The second one states that no extra slot is created, in the sense that any increase in the capacity of a division must come from slots left vacant by preceding divisions.\footnote{ \cite{westkamp2013analysis} introduced the first and second conditions in a simpler matching model without contracts as different properties. \cite{westkamp2013analysis} called the first one "\textit{monotonicity}" and the second one \textit{non-excessive reduction}" property. We adopted these properties to a more general matching model with contracts and combined them in a single property, called monotonicity, to reduce the number of conditions dealt with.}

\subsection*{Conditions on Divisions' Choice Rules}

A choice rule $C_{s}^{k}: 2^{X} \times\mathbb{Z}_{+}\longrightarrow2^{X_{s}}$ satisfies the \textit{substitutes} property if, for all $q\in \mathbb{Z}_{+}$ and for all $x,y\in X$ and $Y\subseteq X \setminus \{x,y\}$,
	\[
	y\notin C_{s}^{k}(Y\cup\{y\};q)\;\Longrightarrow\;y\notin C_{s}^{k}(Y\cup\{x,y\};q).
	\]

 A choice rule $C_{s}^{k}: 2^{X} \times\mathbb{Z}_{+}\longrightarrow2^{X_{s}}$ satisfies \emph{size monotonicity} if, for all $q\in \mathbb{Z}_{+}$ and for all contracts $x\in X$ and sets of contracts $Y\subseteq X$, we have \[
	\mid C_{s}^{k}(Y;q)\mid\leq\mid C_{s}^{k}(Y \cup \{x\};q) \mid.
	\]

The substitutes property and size monotonicity jointly imply the \textit{irrelevance of rejected contracts} (IRC) condition (\cite{aygun2013matching}). A choice rule $C_{s}^{k}$ satisfies the IRC condition if, for any $Y\subseteq X$ and any $x\in X \setminus Y$,  
$$
x\notin C_{s}^{k}(Y\cup \{x\};q) \text{ implies } C_{s}^{k}(Y;q)=C_{s}^{k}(Y\cup \{x\};q).
$$

Both the substitutes property and size monotonicity are standard conditions that impose restrictions on the choice function when the set of alternatives changes, while the capacity remains unchanged. The next property imposes conditions on the choice function when the capacity increases while the set of alternatives remains unchanged. This property is in the spirit of the expansion property of the choice rules defined in \cite{chambers2017choice}.

\begin{definition}
A choice rule $C_{s}^{k}: 2^{X} \times\mathbb{Z}_{+}\longrightarrow2^{X_{s}}$ satisfies \text{quota monotonicity} (QM) if for any $q,\Delta\in\mathbb{\mathbb{Z}}_{+}$ such that for all $Y\subseteq X$,

\begin{equation}
	C_{s}^{k}(Y,q)\subseteq C_{s}^{k}(Y,q+\Delta),\
\end{equation}
and 
\begin{equation}
    \mid C_{s}^{k}(Y,q+\Delta)\mid-\mid C_{s}^{k}(Y,q)\mid\leq \Delta.
\end{equation}

\end{definition}

Condition (3) says that if there is an increase in capacity, the choice rule selects every contract chosen before the increase in capacity from any given set of contracts. Condition (4) simply states that if the capacity increases by $\Delta$, then the difference between the number of contracts chosen with the increase in capacity and the initial capacity cannot exceed $\Delta$. The first condition is the \textit{expansion} property of \cite{chambers2017choice}. The second condition restricts the magnitude of expansion when capacity is changed.

\begin{definition}
An aggregate choice rule $C_{s}$ is a \textbf{GSq rule} if it can be represented as a pair $\left(\left(C_{s}^{k}\right)_{k\in\mathcal{K}_{s}},q_{s}\right)$ of a collection of division choice rules and transfer policy, where each division's choice rule satisfies the substitutes property, size monotonicity and quota monotonicity, and the transfer policy satisfies monotonicity.
\end{definition}

The family of GSq choice rules is a parametrized family of aggregate choice rules derived from choice functions that satisfy both the substitutes property and size monotonicity. The parameterization comes from two primary sources. The transfer function provides one dimension of parameterization. Different monotonic transfer functions generate distinct aggregate choice rules within this family. 

Moreover, parameterization arises from the expansion of divisions' choice rules while maintaining quota monotonicity. A division may be characterized by a sequence of choice rules, with each rule corresponding to a specific integer capacity. By defining the choice rule of division $k$ as $C_{s}^{k}: 2^{X} \times\mathbb{Z}_{+}\longrightarrow2^{X_{s}}$, we emphasize that \textit{capacity} of division $k$ may vary. In a sense, this definition assigns a series of choice functions to division $k$, one for each given integer capacity. When the capacity of a substitutable and size monotonic choice rule is adjusted, the result is a new choice rule that also satisfies these properties. The quota monotonicity property establishes the precise relationship between these capacity-differentiated choice rules. 

A substitutable and size monotonic choice rule can be "expanded" to another such rule through multiple different ways in the sense that multiple different choice rules may expand the same choice rule that satisfies the subsitutes property and size monotonicity. This fact further enriches the parametric space of the GSq family.

When each division's choice rule is q-responsive, which implies the substitutes property, size monotonicity, and QM, the aggregate choice rule becomes a \textbf{dynamic reserves choice rule} introduced in \cite{aygun2020dynamic}. Therefore, the family of GSq rules nests the family of dynamic reserves choice rules. An important example of a choice rule that satisfies the substitute property, size monotonicity, and QM but violates responsiveness is the \textit{ meritorious horizontal choice rule} introduced in \cite{sonmez2022affirmative}. The GSq family also nests the family of slot-specific priorities choice rules, introduced by \cite{kominers2016matching}, because \cite{aygun2020dynamic} shows that each slot-specific priorities choice rule can be represented as a dynamic reserves choice rule.

The aggregate choice rule in the GSq family may fail to satisfy the substitute property and size monotonicity, even though the divisions' choice rules satisfy them. However, a stable and strategy-proof mechanism design is still possible.

\section{Stable and strategy-proof mechanisms} \label{MecDesign}

\cite{hatfield2019hidden} introduce conditions on the choice rules of the institutions to guarantee a stable and strategy-proof matching mechanism.

\begin{definition}[\cite{hatfield2019hidden}]
    A \textbf{completion} of a choice function $C_{s}$ is a choice function $\overline{C}_{s}$, such that for
all $Y\subseteq X$, either $\overline{C}_{s}(Y)=C_{s}(Y)$
or there exists a distinct $x,x^{'}\in\overline{C}_{s}(Y)$ such that $\mathbf{i}(x)=\mathbf{i}(x^{'})$.  

If a choice function $C_{s}$ has a completion that satisfies the substitute property and the IRC condition, then $C_{s}$ is said to be \textbf{substitutably completable}. If every choice rule in a profile $C=(C_{s})_{s\in S}$ is substitutably completable, then $C$ is substitutably completable.
\end{definition}

Let $C_{s}(\cdot;q_{s})$ be an GSq choice rule induced by the transfer function $q_{s}$. We define a related choice function $\overline{C}_{s}(\cdot;q_{s})$ as follows: 

Given a set of contracts $Y\equiv Y^{1}\subseteq X$, a capacity $\overline{q}_{s}$ for institution $s$, and a capacity $\overline{q}^{1}_{s}$ for division 1, we compute the set of chosen contracts $\overline{C}_{s}(Y;q_{s})$ in steps $K_{s}$, where divisions choose in sequential order.
	
\paragraph{Step 1}
Given $Y^{1}$ and $\overline{q}^{1}_{s}$, division 1 chooses $C^{1}_{s}(Y^{1};\overline{q}^{1}_{s})$. Let $r_{1}=\overline{q}^{1}_{s}-\mid C^{1}_{s}(Y^{1};\overline{q}^{1}_{s})\mid$. Let $Y^{2}\equiv Y^{1}\setminus C^{1}_{s}\left(Y^{1};\overline{q}^{1}_{s}\right)$. 
	
\paragraph{Step k. ($2\protect\leq k\protect\leq K_{s})$}
Given $Y^{k}$ and its capacity $q^{k}_{s}(r_{1},...,r_{k-1})$, division $k$ chooses $C^{k}_{s}(Y^{k};q^{k}_{s}(r_{1},...,r_{k-1}))$.
	Let $r_{k}=q^{k}_{s}(r_{1},...,r_{k-1})-\mid C^{k}_{s}(Y^{k};q^{k}_{s}(r_{1},...,r_{k-1}))\mid$.
	Let $Y^{k+1}\equiv Y^{k}\setminus C^{k}_{s}\left(Y^{k};q^{k}_{s}\left(r_{1},\ldots,r_{k-1}\right)\right)$. 
	
The union of the' chosen sets of the divisions is "the" institution's chosen set. That is,
\[
	\overline{C}_{s}(Y;q_{s})\equiv C^{1}_{s}(Y^{1};\overline{q}^{1}_{s})\cup \bigcup_{k=2}^{K_{s}} {C^{k}_{s} (Y^{k};q^{k}_{s}(r_{1},...,r_{k-1}))}.
\]
	
The difference between $C_{s}$ and $\overline{C}_{s}$ is as follows: In the computation of $C_{s}$, if a contract of an individual is chosen by some division, then her other contracts are removed for the rest of the procedure. According to $\overline{C}_{s}$, if an individual's contract is chosen by division $k$, then her other contracts will still be available for the remaining divisions. Note that the completion uses the same divisional choice rules, i.e. $\bar{C}_s^k = C_s^k$, for all $k=1,...,K_s$. 

We are now ready to state our main result.

\begin{theorem}\label{Theorem 1}
    	$\overline{C}_{s}$ is a completion of $C_{s}$. Moreover,  $\overline{C}_{s}$ satisfies the IRC condition, the substitute property, and the size monotonicity.
\end{theorem}

Cumulative offer mechanisms (COMs) have received special attention in the literature. In an COM, individuals propose contracts according to a strict ordering $\triangleright$ of the elements of $X$. In each step, some individual who does not currently have a contract held by any institution proposes his most preferred contract that has not yet been proposed. Then, each institution chooses its most preferred set of contracts according to its choice rule and holds this set until the next step. When multiple individuals are able to propose in the same step, the individual who proposes is determined by ordering $\triangleright$. 

Given the aggregate choice rules of the institutions and the preferences of individuals, the outcome of the COM is calculated by the cumulative offer process (COP) as follows: 

\paragraph{Step 1}
Some individual $i^{1}\in\ I$ proposes her most preferred contract, $x^{1}\in\ X_{i^{1}}$. Institution $\mathbf{s}\left(x^{1}\right)$
holds $x^{1}$ if $x^{1}\in C_{\mathbf{s}\left(x^{1}\right)}\left(\left\{ x^{1}\right\} \right)$,
and rejects $x^{1}$ otherwise. Set $A_{\mathbf{s}\left(x^{1}\right)}^{2}=\left\{ x^{1}\right\} $,
and set $A_{s'}^{2}=\emptyset$ for each $s'\neq\mathbf{s}\left(x^{1}\right)$, i.e., the sets of contracts available to institutions at Step 2. 

\paragraph{Step l}
Some individual $i^{l}\in\ I$, for whom no institution currently holds a contract, proposes her most preferred contact that has not yet been rejected $x^{l}\in X_{i^{l}}\setminus \left(\bigcup_{j \in S} A_j^l\right)$. Institution $\mathbf{s}\left(x^{l}\right)$
holds the set of contracts in $C_{\mathbf{s}\left(x^{l}\right)}\left(A_{\mathbf{s}\left(x^{l}\right)}^{l}\cup\left\{ x^{l}\right\} \right)$
and rejects all other contracts in $A_{s\left(x^{l}\right)}^{l}\cup\left\{ x^{l}\right\} $;
institutions $s'\neq\mathbf{s}\left(x^{l}\right)$ continue to hold
all contracts they held at the end of Step $l-$1. Set $A_{\mathbf{s}\left(x^{l}\right)}^{l+1}=A_{\mathbf{s}\left(x^{l}\right)}^{l}\cup\left\{ x^{l}\right\} $,
and set $A_{s'}^{l+1}=A_{s'}^{l}$ for each $s'\neq\mathbf{s}\left(x^{l}\right)$. 

The algorithm terminates when no individual can propose. Each institution is assigned the set of contracts that it holds in the last step.

The order of contracts offered during the COM does not matter. The outcome of the COM is independent of the order of proposals (\cite{hatfield2017restud}).\footnote{In their Proposition 3, \cite{hatfield2017restud} show that observable substitutability of institutions' choice rules is sufficient for the order independence of the COM. Choice rules in the GSq family satisfy observable substitutability.}

Given $P=\left(P_{i}\right)_{i\in I}$ and $C=\left(C_{s}\right)_{s\in S}$, let $\Phi\left(P,C\right)$ be the result of the COM. Let $\Phi_{i}\left(P,C\right)$ and $\Phi_{s}\left(P,C\right)$ denote the assignment of individual $i$ and institution $s$, respectively. 

\cite{hatfield2017restud} show that when institutions' choice rules satisfy three conditions --observable substituablity, observable size monotonicity, and not manipulability via contractual terms -- the COM is the unique mechanism that is stable and strategy-proof. Moreover, \cite{hatfield2017restud} show that choice rules that admit completions that satisfy the substitute property and size monotonicity satisfy these three conditions. Therefore, combining these results with our Theorem 1, we obtain the following important corollary.

\begin{corollary}
    The COM is the unique mechanism that is stable and strategy-proof when institutions' choice rules are in the GSq family. 
\end{corollary}

\section{Practical Choice Rules in the GSq Family}\label{Applications}

Policy goals in many real-world matching markets naturally give rise to choice rules that are in the GSq family. In this section, we present three applications of real-world resource allocation problems where the proposed choice rules that emerge from policy desiderata are in the GSq family. Namely, we analyze (i) the aggregate choice rule in which the meritorious horizontal choice rule selects candidates in each category in allocation government jobs in India (\cite{sonmez2022affirmative}) and (ii) the simultaneous open-reserve open-choice rules proposed for high school admissions in China (\cite{hu2025verifiable}).

\subsection{Affirmative Action in India}\label{India}

India has implemented a complex affirmative action program in allocating government jobs and admissions to public universities since the 1950s. India's highly regulated reserve system consists of two different provisions: \textit{vertical} and \textit{horizontal}.\footnote{The terms vertical and horizontal first appeared in the Supreme Court judgment in Indra Sawhney vs. Union of India (1992). The case is available at https://indiankanoon.org/doc/1363234/.}

\paragraph{Vertical Reservations}
The reservations for Scheduled Castes (SC), Scheduled Tribes (ST), Other Backward Classes (OBC) and Economically Weaker Sections (EWS) are called vertical reservations. 15\%, 7.5\%, 27\%, and 10\% of the positions available at institutions are designated for these respective groups.\footnote{In 2019, the Indian legislative body instituted a 10\% vertical reservation for a subset of the general category, specifically the Economically Weaker Section (EWS), defined by an annual income threshold below Rs. 8 lakhs.} Individuals not included in these vertical categories are classified as members of the \textbf{general} category (g). The residual 40.5\% of available positions is referred to as \textbf{open category} positions. Any open category positions secured by individuals belonging to SC, ST, OBC, or EWS are not to be deducted from their respective vertical reservations by law. Failure to declare membership in the SC, ST, OBC, or EWS categories relegates candidates to category-g by default. Importantly, disclosure of vertical category affiliation is discretionary.

Let $R=\{SC,ST,OBC, EWS\}$ be the set of reserved categories. In each institution $s\in S$, $\overline{q}_{s}^{SC}$,
$\overline{q}_{s}^{ST}$, $\overline{q}_{s}^{OBC}$, and $\overline{q}_{s}^{EWS}$ positions are earmarked for the categories SC, ST, OBC, and EWS, respectively. The remaining $\overline{q}_{s}^{o}=\overline{q}_{s}-\left(\overline{q}_{s}^{SC}+\overline{q}_{s}^{ST}+\overline{q}_{s}^{OBC}+\overline{q}_{s}^{EWS}\right)$
positions are the open category positions. We denote the set of vertical categories $V=\{o,SC,ST,OBC,EWS\}$, where $o$  denotes the open category. Let $$\overline{q}_{s}=(\overline{q}_{s}^{o},\overline{q}_{s}^{SC},\overline{q}_{s}^{ST},\overline{q}_{s}^{OBC},\overline{q}_{s}^{EWS})$$ be the vector of vertical reservations at institution $s$. 

The function $t:I\rightarrow R\cup\left\{ g\right\} $ denotes individuals' category membership. For every individual $i\in I$, $t(i)$ denotes the category to which individual $i$ belongs. We denote a reserved category membership profile by $T=\left(t(i)\right)_{i\in I}$, and let $\mathcal{T}$ be the set of all possible reserved category membership profiles. 

\paragraph{Horizontal Reservations}
Horizontal reservations are provided to other marginalized groups, such as women and individuals with disabilities\footnote{For an in-depth analysis, see \cite{sonmez2022affirmative}.}. The SCI's judgment in Indra Sawhney (1992) mandates that horizontal reservations must be implemented within each vertical category on a minimum guarantee basis. That is, in each vertical category, horizontally reserved positions must be allocated before any unreserved positions are allocated. However, the SCI judgments have left it ambiguous whether an individual eligible for multiple types of horizontal reservation is counted against a single horizontal category or multiple categories concurrently.

Let $H=\{h_{1},\ldots,h_{L}\}$ be the set of horizontal types.
The correspondence $\rho:\:I\rightrightarrows H\cup\{G\}$ represents the horizontal types of individuals. $\rho(i)\subseteq H\cup\{G\}$ denotes the set of horizontal types that the individual $i$ can claim. $\rho(i)=G$ means that $i$ has no horizontal type.  We denote by $\kappa_{v}^{j}\in \mathbb{Z}_{\geq 0}$ the number of positions reserved for horizontal type $h_{j}\in H$ in the vertical category $v\in V$.
The vector $\kappa_{v}^{s}\equiv(\kappa_{v}^{1},...,\kappa_{v}^{L})\in \mathbb{Z}_{\geq 0}^{L}$
denotes the horizontal reservations in the vertical category $v\in V$
in the institution $s$. Let $\Gamma{s}\equiv\left\{ \kappa_{v}^{s}\right\} _{v\in V}\in \mathbb{Z}_{\geq 0}^{L \times |V|}$ denote the horizontal reservations of the institution $s$. 

\paragraph{De-reservations}
SC and ST are classified as hard reserves, which means that vacant positions reserved for these groups cannot be transferred to other categories. Although reservations for OBC are treated as hard reserves in the context of governmental job allocations, they were suggested and have been implemented as soft reserves in admissions to public colleges, pursuant to the landmark SCI decision in Ashoka Kumar Thakur vs. Union of India (2008).\footnote{The judgment is available at https://indiankanoon.org/doc/1219385/.} The SCI rulings did not provide an explicit procedural framework for de-reserving unfilled OBC positions. The lack of clear rules has led to various inconsistent practices in school admissions in India (\cite{aygun2022dereserve}).

\paragraph{Priority Orders}
Each institution $s$ has a strict priority order $\succ_{s}$ over $\mathcal{I}\cup\{\emptyset_{s}\}$. We write $i\succ_{s}j$ to mean that the applicant $i$ has a higher priority than $j$ at $s$.
Similarly, we write $i\succ_{s}\emptyset_{s}$ to say that applicant $i$ is acceptable
for $s$. $\emptyset_{s}\succ_{s}i$ means that the applicant
$i$ is unacceptable for $s$. The profile of institutions' priorities
is denoted $\succ=(\succ_{s_{1}},...,\succ_{s_{m}})$. For each institution
$s\in\mathcal{S}$, the merit ordering for individuals of type $r\in\mathcal{R}$,
denoted by $\succ_{s}^{r}$, is obtained from $\succ_{s}$ in a straightforward
manner as follows: 
\begin{itemize}
	\item for $i,j\in\mathcal{I}$ such that $t_{i}=r$, $t_{j}\neq r$, $i\succ_{s}\emptyset_{s}$,
	and $j\succ_{s}\emptyset_{s}$, we have $i\succ_{s}^{r}\emptyset_{s}\succ_{s}^{r}j$,
	where $\emptyset_{s}\succ_{t}^{r}j$ means individual $j$ is unacceptable
	for category $r$ at institution $s$. 
	\item for any other $i,j\in\mathcal{I}$, $i\succ_{s}^{r}j$ if and only
if $i\succ_{s}j$. 
\end{itemize}

\paragraph{Individuals' Preferences}
We define $\mathcal{X}\equiv\underset{t(i)=g}{\bigcup}\{i\}\times S \times\{o\}\underset{t(i)\in R}{\bigcup}\{i\}\times S \times\{t(i),o\}$
as the set of all contracts. Each contract $x\in\mathcal{X}$ is between
an individual $\mathbf{i}(x)$ and an institution $\mathbf{s}(x)$
and specifies a vertical category $\mathbf{t}(x)$ in which the individual
$\mathbf{i}(x)$ is admitted. 

Each individual $i$ with $t(i)=g$ has a preference for $\mathcal{S\times}\left\{ o\right\} \cup\left\{ \emptyset_{i}\right\} $. Each individual $i$ with $t(i)=r\in R$ has a preference over $\mathcal{S}\times\left\{ r,o\right\} \cup\left\{ \emptyset_{i}\right\} $,
where $\emptyset_{i}$ denotes the outside option for individual $i$.\footnote{We assume that individuals do not have preferences for horizontal types as there is no evidence that individuals care about the horizontal types under which they are admitted. All discussion on reservation policy in India centers on vertical reservations.} We write $(s,v)P_{i}(s^{'},v^{'})$ to mean that the individual $i$ strictly prefers admission to institution $s$ through the vertical category $v$ to admission to institution $s^{'}$ via vertical category $v^{'}$. The \emph{at-least-as-well} relation $R_{i}$ is obtained from $P_{i}$ as follows: $(s,v)R_{i}(s^{'},v^{'})$ if and only if $(s,v)P_{i}(s^{'},v^{'})$ or $(s,v)=(s^{'},v^{'})$.
An institution and vertical category pair $(s,v)$ is \emph{acceptable}
to individual $i$ if it is at least as good as the outside option
$\emptyset_{i}$ and is \emph{unacceptable} to her if it is worse than
the outside option $\emptyset_{i}$. 

\subsection*{Meritorious Horizontal Choice Rule}
We consider \textit{one-to-one} horizontal matching, where an admitted applicant with multiple eligible horizontal types is counted against \textit{only one} of these types. \cite{sonmez2022affirmative} proposed the \textit{meritorious horizontal} choice rule, denoted $C^{m}$, in this setting. 
Within each vertical category $v$, the authors consider the assignment of horizontally reserved positions $H^{v}=\bigcup_{j=1}^{L} H_{j}^{v}$ to a set of individuals $I$ as a one-to-one matching problem, where $H_{j}^{v}$ denotes the set of positions reserved for horizontal type $h_{j}\in H$ in vertical category $v$.  The number of positions in $H_{j}^{v}$ is denoted by $\kappa_{v}^{j}$.

Given a vertical category $v\in\mathcal{V}$ and a set of individuals $I$ such that $t(i)=v$ for all $i\in I$, define the one-to-one matching between the individuals and horizontally reserved positions as a mapping $\mu:I\rightarrow H^{v}\bigcup\left\{ \emptyset\right\} $ such that for any $i\in I$ and $h_{j}\in H$, 
\begin{itemize}
	\item $\mu(i)\in H_{j}^{v}$ implies $h_{j}\in\rho(i)$, and for any $i,k\in I$, 
	\item $\mu(i)=\mu(k)\neq\emptyset$ implies $i=k$.
\end{itemize}

Let $n^{v}(I)$ represent the maximum number of horizontally reserved positions within the vertical category $v$. The matching between individuals and horizontally reserved positions is said to have ``\textit{maximum cardinality}" if there is no alternative assignment that would strictly increase the total number of positions allocated to eligible individuals. For a given vertical category $v$ and a set of individuals $I$ belonging to category $v$, an individual $i$ is said to ``\textit{increase horizontal reservation utilization}" if $n^{v}(I\cup\{i\})=n^{v}(I)+1$.

Given a vertical category $v$, a set of applicants $A$ such that $t(i)=v$ for all $i\in A$, a profile of the horizontal types of applicants $\left(\rho(i)\right)_{i\in A}$, the meritorious horizontal choice rule $C^{m}$ selects the applicants as follows: 

\paragraph{$\mathbf{Step\;1.1}$ }
\emph{Choose the highest merit score individual in $A$ with a horizontal
	type. Denote this individual by $i_{1}$ and let $A_{1}=\left\{ i_{1}\right\} $.
	If no such individual exists, proceed to Step 2. }

\paragraph{$\mathbf{Step\;1.n}$}
\emph{Choose the highest merit score individual in $A\setminus A_{n-1}$
	who increase the horizontal reservation utilization given $A_{n-1}$,
	if any. Denote this individual by $i_{n}$ and let $A_{n}=A_{n-1}\cup\{i_{n}\}$.
	If no such individual exists, proceed to Step 2. }

\paragraph{$\mathbf{Step\;2}$ }
\emph{Fill remaining positions following the merit score with individuals
	among the unassigned individuals until all positions are assigned
	or all individuals are selected. }

\cite{fleiner2001matroid} showed that the first stage of $C^{m}$ satisfies the substitute property. \cite{sonmez2022affirmative} proved that $C^{m}$ satisfies size monotonicity. We prove that $C^{m}$ also satisfies the quota monotonicity. Thus, any aggregate choice rule in which the divisions select according to $C^{m}$ under a monotone transfer policy is in the GSq family.

\subsection*{Aggregate Choice Rule with De-reservation}

According to the aggregate choice rule with transfer, $C^{mT}$, vacant OBC positions are reverted to an open category and filled following merit scores. The rule $C^{mT}$ can be written as
\[
C_{s}^{mT}\equiv\left((\left(C_{v}^{m}\right)_{v\in\mathcal{V}},C_{D}^{m}),q_{s}\right),
\]

where 
\begin{itemize}
	\item the precedence sequence is $o-SC-ST-OBC-EWS-D$, where $D$ denotes
	``\emph{de-reserved positions}'' that are reverted from OBC,
	and 
	\item $q_{s}=\left(q_{s}^{o},q_{s}^{SC},q_{s}^{ST},q_{s}^{OBC},q_{s}^{EWS},q_{s}^{D}\right)$
	such that $q_{s}^{v}=\overline{q}_{s}^{v}$ for all $v\in\left\{ o,SC,ST,OBC,EWS\right\} $,
	and $q_{s}^{D}=r_{OBC}$ is the number of vacant OBC positions, and
    \item $C_{C}^{m}$ is a $q_{s}^{D}$-responsive choice rule induced by the merit ranking of individuals.
\end{itemize}

Our next result states that when each vertical category $v\in\mathcal{V}$ chooses applicants through $C_{v}^{m}$, under any monotonic transfer policy, the aggregate choice rule $C_{s}^{mT}$ is in the GSq family.

\begin{proposition}\label{Prop1}
    $C_{s}^{mT}$ belongs to the GSq family.
\end{proposition}

An immediate corollary of Proposition \ref{Prop1} is as follows.

\begin{corollary}
    Let $\Phi^{mT}$ be the COM under the profile of choice rules $ C^{mT}=\left(C_{s}^{mT}\right)_{s\in\mathcal{S}}$. Then $\Phi^{mT}$ is stable and strategy-proof for individuals. 
\end{corollary}

\subsection{Affirmative Action in Chinese High School Admissions}

In China, public high schools are subject to an affirmative action policy in which high schools reserve a proportion of their available seats for graduates of low-performing middle schools. High school admissions are merit-based and highly competitive. The affirmative action policy requires high schools to reserve 50\% or more of their seats exogenously to middle schools. \cite{hu2025verifiable} study a novel aspect of the Chinese affirmative action policy, called ``\textit{verifiability}," which has been implemented in almost every city in China.\footnote{Verifiability of an allocation mechanism ensures that students can verify their assignments using only their private information (e.g., their preferences over schools) and publicly observed cutoff scores; It is an important aspect of the transparency and accountability of the admission process.} The authors analyze several choice rules that are used in Chinese cities. They also propose new choice rules. To define them, we first introduce their notation. 

There is a finite set of middle school students $\mathcal{I}$. Let $q_s$ be the capacity of the high school $s$. Let $\mathcal{M}$ be the finite set of middle schools. The high school $s$ reserves $q_{s}^{m}$ seats for students from the middle school $m$. We assume that $\sum_{m\in M}q_{s}^{m}<q_{s}$, where $q_{s}^{o}=q_{s}-\sum_{m\in M}q_{s}^{m}$ denotes the number of open seats in high school $s$. The function $\tau: \mathcal{I} \rightarrow \mathcal{M}$ is such that $\tau(i)$ denotes the middle school student $i$ belongs to. Let $\mathcal{I}^{m}=\tau^{-1}(m)$ and $I^{m}\equiv I\cap \mathcal{I}^{m}$. 

There are two DA-based mechanisms that allocate open and reserved seats simultaneously. In these mechanisms, each high school employs the choice rules $\mathcal{C}^{SimRO}$ and $\mathcal{C}^{SimORO}$ that we define below.

\subsection*{The Simultaneous-reserve-open Choice Rule \texorpdfstring{$\mathcal{C}^{SimRO}$}{C SimRO}}
This choice rule is used in cities such as Guiyan and Tianjin for high school admissions. For any given set $I\subseteq \mathcal{I}$ of students, the choice rule $\mathcal{C}^{SimRO}$ selects the students as follows: 

\paragraph{Step 1}
The reserve seats are allocated for each $m\in \mathcal{M}$ to the highest priority students from $I^{m}$ to capacity $q_{s}^{m}$.

\paragraph{Step 2}
All remaining seats are assigned as open seats to the highest-ranked students among the remaining until all seats are allocated or all students are admitted.

Note that unfilled reserved seats are allocated as open seats in the second stage. 

\begin{proposition}
   $\mathcal{C}^{SimRO}$ is a GSq choice rule.  
\end{proposition}

\subsection*{The Simultaneous-open-reserve-open Choice Rule \texorpdfstring{$\mathcal{C}^{SimORO}$}{C SimORO}}
In Chinese cities Taiyuan, Zengzhou, and Hefei, among many others, for high school admissions, the DA mechanism with the choice rule $\mathcal{C}^{SimORO}$ is used. For any given set $I\subseteq \mathcal{I}$ of students, the choice rule $\mathcal{C}^{SimORO}$ selects the students in three steps. 

\paragraph{Step 1}
Open seats are allocated to students with the highest priority, regardless of their type, up to capacity $q_{s}^{o}$.

\paragraph{Step 2}
The reserve seats are allocated for each $m\in \mathcal{M}$ to the remaining students with the highest priority from $I^{m}$ up to capacity $q_{s}^{m}$.

\paragraph{Step 3}
All remaining seats are assigned as open seats to the highest ranked students among the remaining until all seats are allocated or all students are admitted.

The unfilled seats in the second stage are allocated as open seats in the third stage. 

\begin{proposition}
   $\mathcal{C}^{SimORO}$ is a GSq choice rule.
\end{proposition}

Another version of the DA mechanism used for high school admissions in some Chinese cities differs from the two mechanisms above in that each school is divided into two subschools: \textit{open} and \textit{reserved} subschools. For school $s\in \mathcal{S}$, the open subschool $s^{o}$ has capacity $q_{S}^{o}$, and the reserved subschool $S^{r}_{s}$ has the capacity vector $q_{s}^{r}=(q_{S}^{m})_{m\in \mathcal{M}}$. We instead model it in matching with contracts framework. Students are required to submit strict preferences over subschools. Then, the student-proposing DA coupled with the following choice rules is run to find a matching.\footnote{\cite{pathak2024fair} is the first paper to create a hypothetical matching market between individuals and reserve categories in the context of rationing life-saving medical resources. They define a DA-induced matching as the outcome of the individual-proposing DA for a given "artificial" strict preferences of individuals over categories.} In each round of the DA, each unassigned student applies to the highest-ranked subschool in their preferences. 

We denote a contract between the student $i$ and the school $s$ by $x=(i,s,t)$, where $t\in \{o,r\}$ denotes the seat category under which the student is accepted. Since each student belongs to a middle school in $\mathcal{M}$, each student has two contracts with a given high school. Let $\mathcal{X}=\mathcal{I} \times \mathcal{S}\times \{o,r\}$ denote the set of all contracts. For a given set of contracts $X\in \mathcal{X}$, we let $X_{i}$ and $X_{s}$ denote the set of contracts associated with the student $i$ and the school $s$, respectively. Similarly, let $X_{o}$ and $X_{r}$ denote the contracts with terms $o$ and $r$, respectively. Note that $X_{o} \cup X_{r}=X$.

\subsection*{The Simultaneous-separate Choice Rule \texorpdfstring{$\mathcal{C}^{SimSep}$}{C SimSep}}
Given a set of contracts $X$, the choice rule $\mathcal{C}^{SimSep}$ selects contracts as follows: 

\paragraph{Step 1} 
Consider only the contracts in $X_{o}$. Choose contracts associated with the highest-ranked students up to capacity $q_{s}^{o}$. 

\paragraph{Step 2}
Consider only the contracts in $X_{r}$. For each $m\in  \mathcal{M}$, choose contracts of the highest-ranked students such that $(X_{r})_{i} \in I^{m}$ up to capacity $q_{s}^{m}$.  

This rule does not allocate reserved positions that are left unfilled. Thus, it is wasteful. 

\begin{proposition}
    $\mathcal{C}^{SimSep}$ is a GSq choice rule.
\end{proposition}

\subsection*{The Simultaneous-flexible Choice Rule \texorpdfstring{$\mathcal{C}^{SimFlex}$}{C SimFlex}}
Given a set of contracts $X$, the choice rule $\mathcal{C}^{SimFlex}$ selects contracts as follows: 

\paragraph{Step 1}
Consider only the contracts in $X_{r}$. For each $m\in  \mathcal{M}$, choose contracts of the highest-ranked students such that $(X_{r})_{i} \in I^{m}$ up to capacity $q_{s}^{m}$.

\paragraph{Step 2}
Consider only the contracts in $X_{o}$. Allocate all remaining seats (open seats and leftover reserved seats) by choosing contracts associated with the highest-ranked students.

\begin{proposition}
    $\mathcal{C}^{SimFlex}$ is a GSq choice rule.
\end{proposition}

\subsection*{The Simultaneous-open-reserve Choice Rule \texorpdfstring{$\mathcal{C}^{SimOR}$}{C SimOR}}
Let $I$ be the set of students and $s$ be the only school. Define $X=I\times \{s\}\times \{o,r\}$. Assume that each student $i\in I$ ranks the contract with the open seat over the contract with the reserved seat. Run the student proposing DA while the school $s$ selects students using $\mathcal{C}^{SimFlex}$. The result of the DA procedure defines $\mathcal{C}^{SimOR}(I)$.

\cite{hu2025verifiable} show that $\mathcal{C}^{SimOR}$ does not belong to the class of slot-specific priorities choice rules of \cite{kominers2016matching}. 

\begin{proposition}
    $\mathcal{C}^{SimOR}$ is a GSq choice rule.
\end{proposition}

\section{Conclusion}\label{sec:concl}

This paper introduced a new family of choice rules for institutions with multiple divisions. The GSq framework unifies and extends various choice rules that appear naturally in real-world market design applications, particularly those involving complex affirmative action policies. We showed that when institutions use GSq choice rules, the Cumulative Offer Mechanism (COM) is the unique mechanism that satisfies both stability and strategy-proofness. This uniqueness result provides strong theoretical support for using COM in settings with complex distributional constraints.

We demonstrated that prominent choice rules used in practice, such as aggregate choice rules that use meritorious horizontal choice rules in each vertical category in India and several choice rules used in high school admissions in numerous cities in China, are members of the GSq family. We provide a unified theoretical framework for analyzing them. 

The framework developed in this paper provides a foundation for analyzing complex matching markets with distributional constraints. By unifying various existing approaches under the GSq family, it opens new avenues for both theoretical analysis and practical market design.

\section{Appendix}

\section*{Proof of Theorem 1}

We first show that $\overline{C}_{s}$ is a completion of $C_{s}$.

Consider an offer set $Y=Y^{1}\subseteq X$ and assume that there is no pair of contracts $z,z^{'}\in Y^{1}$ such that $i(z)=i(z^{'})$ and $z,z^{'}\in\overline{C}_{s}(Y;q_{s})$. We want to show that $\overline{C}_{s}(Y;q_{s})=C_{s}(Y;q_{s}).$ 
		
Let $(Y_{j},r_{j},Y^{j+1})$ and $(Z_{j},\overline{r}_{j},Z^{j+1})$ be sequences of a set of contracts chosen by division $j$, the number of vacant slots in division $j$, and the set of contracts that remains in the choice procedure after division $j$ selects under $C_{s}$ and $\overline{C}_{s}$, respectively. 
		
Given $\bar{q}_{s}^{1}$ and $Y^{1}=Z^{1}$, we have $Z_{1}=\overline{C}^{1}_{s}(Z^{1};\overline{q}_{1}^{s})=C^{1}_{s}(Y^{1};\bar{q}_{s}^{1})=Y_{1}$. Moreover, $\overline{r}_{1}=r_{1}$ and $\overline{q}_{2}^{s}(\overline{r}_{1})=q_{2}^{s}(r_{1})$. Suppose that for all $j\in\{2,...,k-1\}$, we have $Y_{j}=Z_{j}$. We need to show that $Y_{k}=Z_{k}$. 
		
Since $Y_{j}=Z_{j}$ for all $j=1,\ldots,k-1$, we obtain $r_{j}=\overline{r}_{j}$ for all $j=1,\ldots,k-1$, which implies by the definition of the transfer function $$q^{k}_{s}(r_{1},...,r_{k-1})=\overline{q}_{s}^{k}(\overline{r}_{1},...,\overline{r}_{k-1}).$$ 

Since there are no two contracts of an individual chosen by $\overline{C}_{s}$, all the remaining contracts of individuals---whose contracts were chosen by previous divisions---are rejected by $C^{k}_{s}(\overline{Y}^{k};\overline{q}_{s}^{k}(\overline{r}_{1},...,\overline{r}_{k-1}))$. Therefore, the IRC of $C^{k}_{s}$ implies that\footnote{Divisions' choice rules are assumed to satisfy the substitutes property and size monotonicity. The conjugation of the two implies the irrelevance of rejected contracts (IRC) condition.} $$C^{k}_{s}(\overline{Y}^{k};\overline{q}_{k}^{s}(\overline{r}_{1},...,\overline{r}_{k-1}))=C_{s}^{k}(Y^{k};q^{k}_{s}(r_{1},...,r_{k-1})).$$ 

Thus, we obtain $\overline{Y}_{k}=Y_{k}$, $\overline{r}_{k}=r_{k}$, and $\overline{q}^{k+1}_{s}(\overline{r}_{1},...,\overline{r}_{k})=q^{k+1}_{s}(r_{1},...,r_{k})$. Since each division selects the same set of contracts under $C_{s}$ and $\overline{C}_{s}$, the result follows. 
	
In the following, we prove that the rule $\overline{C}_{s}$ satisfies the irrelevance of rejected contracts (IRC) condition, substitutes property, and size monotonicity.

\subsection*{IRC Condition} 
Consider $Y=Y^{1}\subseteq X$ such that $Y\neq\overline{C}_{s}(Y;q_{s})$. Take $x\in Y\setminus\overline{C}_{s}(Y;q_{s})$. We need to show that $\overline{C}_{s}(Y;q_{s})=\overline{C}_{s}(Y\setminus\{x\};q_{s}).$
		
Let $Z^{1}=Y^{1}\setminus\{x\}$. Let $(Y_{j},r_{j},Y^{j+1})$ and $(Z_{j},\tilde{r}_{j},Z^{j+1})$ be sequences of the set of chosen contracts, the number of vacant slots, and the remaining set of contracts for division $j=1,...,K_{s}$ under $\overline{C}_{s}$ from $Y$ and $Z$, respectively. 
		
Since the choice function of division 1 satisfies the IRC, we have $Y_{1}=Z_{1}$. Moreover, $r_{1}=\tilde{r}_{1}$ and $Y^{2}\setminus\{x\}=Z^{2}$. By induction, for each $j=2,...,k-1$, assume that
\[
		Y_{j}=Z_{j},\;r_{j}=\tilde{r}_{j},\;\text{and} \;Y^{j}\setminus\{x\}=Z^{j}.
\]
		
We need to show that the above equalities hold for $j=k$. Since $x\notin\overline{C}_{s}(Y;q_{s})$ and divisions' choice functions satisfy the IRC condition, we obtain 
\[
		\overline{C}^{k}_{s}(Y^{k};q^{k}_{s}(\overline{r}_{1},...,\overline{r}_{k-1}))=\overline{C}^{k}_{s}(Z^{k};q^{k}_{s}(\tilde{r}_{1},...,\tilde{r}_{k-1})).
\]

By our inductive assumption that $r_{j}=\tilde{r}_{j}$ for each $j=2,...,k-1$, division $k$'s capacity is the same under both choice processes. The number of remaining slots is the same as well. That is, $r_{k}=\tilde{r}_{k}$.

Moreover, we know that $x$ is not chosen from the set $Z^{k}\cup\{x\}$, then we have $$Y^{k+1}=Z^{k+1}\cup\{x\}.$$ 
		
Since for all $j\in\{1,...,K_{s}\}$, $Y_{j}=Z_{j}$, we conclude $\overline{C}_{s}(Y;q_{s})=\overline{C}_{s}(Z;q_{s})$. 
	
\subsection*{Substitutes Property}
Consider $Y=Y^{0}\subseteq X$ such that $Y\neq\overline{C}_{s}(Y;q_{s})$. Let $x$ be such that $x\in Y\setminus\overline{C}_{s}(Y;q_{s})$, and let $z$ be an arbitrary contract in $X\setminus Y$. Let $Z=Z^{0}=Y\cup\{z\}$. We need to show that $x\notin\overline{C}_{s}(Z;q_{s}).$ 
		
Let $(Y_{j},r_{j},Y^{j})$ and $(Z_{j},\tilde{r}_{j},Z^{j})$ be sequences of the set chosen contracts, the number of vacant slots, and the set of remaining contracts of the divisions $j=1,...,K_{s}$ from $Y$ and $Z$, respectively, under $\overline{C}_{s}$. 
		
If $z\notin \overline{C}_{s}(Z;q_{s})$, then the IRC of $\overline{C}_{s}$ implies $\overline{C}_{s}(Z;q_{s})=\overline{C}_{s}(Y;q_{s})$. Thus, $x\notin\overline{C}_{s}(Z;q_{s})$. 
		
Now suppose $z\in\overline{C}_{s}(Z;q_{s})$. Let $k^*$ be the division in which $z$ is chosen. 

Since $z\notin Z_{j}$, by the IRC of the choice rules, we have $Z_{j}=Y_{j}$ for $j=1,...,k^{*}-1$. Thus, we obtain $r_{j}=\tilde{r}_{j}$, for $j=1,...,k^{*}-1$. Moreover, for each $j=1,...,k^{*}-1$, the division capacities are the same under the two choice processes, starting with $Y$ and $Z$, respectively. That is, $$q_{s}^{j}(r_1,...,r_{j-1})=q_{s}^{j}(\widetilde{r}_{1},...,\widetilde{r}_{j-1}).$$

Now, we consider division $k^{*}$. We know $z\in Z_{k^*}=C_{s}^{k^*}(Z^{k^*-1};q_s^{k^*}(\widetilde{r}_1,...,\widetilde{r}_{k^*-1}))$.

By the substitutes property of $C_s^{k^*}$, all contracts that are rejected from $Y^{k^*-1}$ are rejected from $Z^{k^*-1}=Y^{k^*-1} \cup \{z\}$. By size monotonicity, $\mid Z_{k^*}\mid\geq\mid Y_{k^*}\mid$. There are two possibilities:

\begin{enumerate}
    \item $Z_{k^*}=Y_{k^*}\cup\{z\}, \text{or}$
    \item $Z_{k^*}=Y_{k^*}\cup\{z\}\setminus\{w\} \text{for some } w\in Y_{k^*}$.
\end{enumerate}

In the former case, we obtain $Y^{k^*}=Z^{k^*}$ and $r_{k^*}=1+\widetilde{r}_{k^*}$. Moreover, by the monotonicity of capacity transfers, we have $$q^{k^{*}+1}_{s}(\widetilde{r}_{1},...,\widetilde{r}_{k^*})\leq q^{2}_{s}(r_{1},...,r_{k^*})\leq1+q^{2}_{s}(\widetilde{r}_{1},...,\widetilde{r}_{k^*}).$$ 

In the latter case, we obtain $Z^{k^*}=Y^{k^*}\cup\{w\}$ and $r_{k^*}=\widetilde{r}_{k^*}$. Since $r_j=\widetilde{r}_j$ for $j=1,...,k^*$, we thus have $$q^{k^*+1}_{s}(r_{1},...,r_{k^*})=q^{k^*+1}_{s}(\widetilde{r}_{1},...,\widetilde{r}_{k*}).$$
		
Suppose now that for all $\gamma=k^{*},...,k-1$, one of the following hold. 
\begin{equation} \label{first}
    \left[Z^{\gamma}=Y^{\gamma}\cup\{w\}\;\text{for some}\;w\;\text{and}\;q^{\gamma+1}_{s}(\tilde{r}_{1},...,\tilde{r}_{\gamma})=q^{\gamma+1}_{s}(r_{1},...,r_{\gamma})\right]
\end{equation}
or 
\begin{equation}\label{second}
    \left[Z^{\gamma}=Y^{\gamma}\;\text{and}\;q^{\gamma+1}_{s}(\tilde{r}_{1},...,\tilde{r}_{\gamma})\leq q^{\gamma+1}_{s}(r_{1},...,r_{\gamma})\leq1+q^{\gamma+1}_{s}(\tilde{r}_{1},...,\tilde{r}_{\gamma})\right].
\end{equation}

We have already shown that it holds for $\gamma=k^*$. Assuming that one of them holds for $\gamma=k^*,...,k-1$, we need to show that one of them holds for $\gamma=k$. 
		
If (\ref{first}) holds for $\gamma=k-1$,  we have $Z^{k-1}=Y^{k-1}\cup\{w\}$ for some $w$. If $w\notin Z_{k}$, then exactly the same set of contracts is chosen from $Y^{k-1}$ and$Z^{k-1}$. This is because the capacity of the division $k$ is the same in both choice processes and $C_{s}^{k}$ satisfies the IRC condition. We then conclude $Z^{k}=Y^{k}\cup\{w\}$. Moreover, since the number of vacant slots in division $k$ will be the same in both processes, we obtain \[q^{k+1}_{s}(r_{1},...,r_{j})=q^{k+1}_{s}(\tilde{r}_{1},...,\tilde{r}_{j}).\]

If $w\in Z_{k}$, we have two possibilities, depending on whether the capacity of division $k$ is exhausted in the choice process starting with $Y^{0}$ or not. If it is not exhausted (that is, $|Y_{k}|<q_{s}^{k}(r_1,...,r_{k-1})$), then we obtain 
\[
		C_{s}^{k}(Z^{k-1};q^{k}_{s}(\tilde{r}_{1},...,\tilde{r}_{k-1}))=\{w\}\cup C_{s}^{k}(Y^{k-1};q^{k}_{s}(r_{1},...,r_{k-1})),
\]

which implies that $Z^{k}=Y^{k}$. Moreover, we have $r_{k}=\tilde{r}_{k}+1$.

The monotonicity of the transfer function implies the following.
\[
		q_{s}^{k+1}(\tilde{r}_{1},...,\tilde{r}_{k})\leq q_{s}^{k+1}(r_{1},...,r_{k})\leq1+q_{s}^{k+1}(\tilde{r}_{1},...,\tilde{r}_{k}).
\]

Note that the first inequality follows from the first condition of the monotonicity of transfer function and the observation that $\tilde{r}_{i}\leq r_{i}$ for all $i=1,...,k$. The second inequality follows from the second condition of the monotonicity of the transfer function. Note that $|Z_{1}|+\cdots+|Z_{k}|=1+|Y_{1}|+\cdots+|Y_{k}|$.
		
If the capacity of the division $k$ is exhausted (that is, $|Y_{k}|=q_{s}^{k}(r_1,...,r_{k-1})$), then $w\in Z_{k}$ implies that there exists a contract $\upsilon$ that is chosen from $Y^{k-1}$ but rejected from $Z^{k-1}$. Then, we obtain $Z^{k}=Y^{k}\cup\{\upsilon\}$ since $C_{s}^{k}$ satisfies the substitutes property and IRC. 

In addition, the capacity of division $k$ is the same in both processes. In this case, we have $r_{k}=\tilde{r}_{k}=0$. Since $\tilde{r}_{i} = r_{i}$ for all $i=1,...,k$, we obtain, by the definition of the transfer policy, $q^{k+1}_{s}(\tilde{r}_{1},...,\tilde{r}_{k})=q^{k+1}_{s}(r_{1},...,r_{k})$.
		
We now analyze the latter case, in which condition (\ref{second}) holds. That is, we have $Z^{k-1}=Y^{k-1}$ and either $q^{k}_{s}(r_{1},...,r_{k-1})=q^{k}_{s}(\tilde{r}_{1},...,\tilde{r}_{k-1})$ or $q^{k}_{s}(r_{1},...,r_{k-1})=1+q^{k}_{s}(\tilde{r}_{1},...,\tilde{r}_{k-1})$. 

If $q_{s}^{k}(r_{1},...,r_{k-1})=q_{s}^{k}(\tilde{r}_{1},...,\tilde{r}_{k-1})$, then, given that $Z^{k-1}=Y^{k-1}$, we have $Z^{k}=Y^{k}$. This also implies $r_{k}=\tilde{r}_{k}$. Note that given $\tilde{r}_{i}\leq r_{i}$ for all $i=1,...,k$, the first condition of the monotone transfer function implies 

\begin{equation}
    q^{k+1}_{s}(r_{1},...,r_{k})\geq q^{k+1}_{s}(\tilde{r}_{1},...,\tilde{r}_{k}).
\end{equation}

Moreover, from the second condition of monotonicity, we can write

\begin{equation*}
    \sum_{m=2}^{k+1} [q_{s}^{m}(r_{1},...,r_{m-1})-q_{s}^{m}(\widetilde{r}_{1},...,\widetilde{r}_{m-1})]\leq\sum_{m=1}^{k-1}[r_{m}-\widetilde{r}_{m}].
\end{equation*}

We can rewrite the above inequality as follows.

\begin{equation*}
    q^{k+1}_{s}(r_{1},...,r_{k}) + \sum_{m=1}^{k} [q_{s}^{m}(r_{1},...,r_{m-1})-q_{s}^{m}(\widetilde{r}_{1},...,\widetilde{r}_{m-1})]-\sum_{m=1}^{k-1}[r_{m}-\widetilde{r}_{m}]\leq  q^{k+1}_{s}(\tilde{r}_{1},...,\tilde{r}_{k}).
\end{equation*}

Rearranging the terms gives us the following. 

\begin{equation*}
    q^{k+1}_{s}(r_{1},...,r_{k}) + \sum_{m=1}^{k} [q_{s}^{m}(r_{1},...,r_{m-1})-r_{m}]-\sum_{m=1}^{k}[q_{s}^{m}(\widetilde{r}_{1},...,\widetilde{r}_{m-1})-\widetilde{r}_{m}]\leq  q^{k+1}_{s}(\tilde{r}_{1},...,\tilde{r}_{k}).
\end{equation*}

The terms in the summations are the number of contracts in the chosen sets $Y_{m}$ and $Z_{m}$, respectively. Plugging them in gives us the following. 

\begin{equation*}
    q^{k+1}_{s}(r_{1},...,r_{k}) + \sum_{m=1}^{k} |Y_{m}|-\sum_{m=1}^{k}|Z_{m}|\leq  q^{k+1}_{s}(\tilde{r}_{1},...,\tilde{r}_{k}).
\end{equation*}

Note that $\sum_{m=1}^{k} |Y_{m}|-\sum_{m=1}^{k}|Z_{m}|=-1$. Because we have $|Z|=|Y|+1$ and $Z^{k}=Y^{k}$.

Thus, we obtain

\begin{equation}
    q^{k+1}_{s}(r_{1},...,r_{k})\leq 1+ q^{k+1}_{s}(\tilde{r}_{1},...,\tilde{r}_{k}). 
\end{equation}

Inequalities (7) and (8) together imply $$q^{k+1}_{s}(\tilde{r}_{1},...,\tilde{r}_{k}) \leq q^{k+1}_{s}(r_{1},...,r_{k}) \leq 1+ q^{k+1}_{s}(\tilde{r}_{1},...,\tilde{r}_{k}).$$

If $q^{k}_{s}(r_{1},...,r_{k-1})=1+q^{k}_{s}(\tilde{r}_{1},...,\tilde{r}_{k-1})$,
then, given $Z^{k-1}=Y^{k-1}$, we have two cases to consider.
		
\textbf{Case 1:} If $C_{s}^{k}(Z^{k-1};q^{k}_{s}(\tilde{r}_{1},...,\tilde{r}_{k-1}))=C_{s}^{k}(Y^{k-1};q^{k}_{s}(r_{1},...,r_{k-1})),$ then we have $Z^{k}=Y^{k}$. In addition, the monotonicity of the transfer function implies
\[
		q^{k+1}_{s}(\tilde{r}_{1},...,\tilde{r}_{k})\leq q^{k+1}_{s}(r_{1},...,r_{k})\leq1+q^{k+1}_{s}(\tilde{r}_{1},...,\tilde{r}_{k}).
\]
		
\textbf{Case 2:} If $C_{s}^{k}(Z^{k-1};q^{k}_{s}(\tilde{r}_{1},...,\tilde{r}_{k-1}))\cup\{\vartheta\}=C_{s}^{k}(Y^{k-1},q^{k}_{s}(r_{1},...,r_{k-1}))$ for some $\vartheta$, then we have $Z^{k}=Y^{k}\cup\{\vartheta\}$. Moreover, the monotonicity of the transfer function implies
\[
		q_{s}^{k+1}(r_{1},...,r_{k})=q_{s}^{k+1}(\tilde{r}_{1},...,\tilde{r}_{k}).
\]
		
To conclude the proof, since either (\ref{first}) or (\ref{second}) holds in each division, given that $x\notin Y_{k}$, we have $x\notin Z_{k}$ for all $k=1,...,K_{s}$. We conclude that $x\notin\overline{C}_{s}(Y\cup\{z\};q_{s})$. Thus, $\overline{C}_{s}$ satisfies the substitutes property.

\subsection*{Size Monotonicity}
Take $Y$ and $Z$ such that $Y\subseteq Z\subseteq X$. Let $q_{s}$
be the transfer function. We want to show that\[
		\mid\overline{C}_{s}(Y;q_{s})\mid\leq\mid\overline{C}_{s}(Z;q_{s})\mid. \] 
        
Let $(Y_{j},r_{j},Y^{j})$ and $(Z_{j},\tilde{r}_{j},Z^{j})$ be the
sequences of the set of chosen contracts, the numbers of unfilled slots, and the sets of remaining contracts for divisions $j=1,...,K_{s}$ in choice processes starting with $Y=Y^{0}$ and $Z=Z^{0}$, respectively. 
		
For the first division with capacity $\overline{q}_{s}^{1}$, since the choice function is size monotonic, we have \[
		\mid Y_{1}\mid=\mid C^{1}_{s}(Y^{0};\overline{q}_{s}^{1})\mid\leq\mid C^{1}_{s}(Z^{0};\overline{q}_{s}^{1})\mid=\mid Z_{1}\mid.
		\] 
        
This, in turn, implies $r_{1}=\overline{q}_{s}^{1}-\mid Y_{1}\mid\geq\tilde{r}_{1}=\overline{q}_{s}^{1}-\mid Z_{1}\mid$. Moreover, we have $Y^{1}\subseteq Z^{1}$. To see this, consider $y\in Y^{1}$. This means that $y\notin Y_{1}$. If $y$ is not chosen
from $Y^{0}$, then it cannot be chosen from a larger set $Z^{0}$ due to the substitutes property.
		
Suppose that $\tilde{r}_{j}\leq r_{j}$ and $Y^{j}\subseteq Z^{j}$ hold for all $j=1,...,k-1$. We need to show that 
\begin{enumerate}
    \item $\tilde{r}_{k}\leq r_{k}$, and
    \item $Y^{k}\subseteq Z^{k}$. 
\end{enumerate}

We first show (1). Given that $\tilde{r}_{j}\leq r_{j}$ for all $j=1,...,k-1$,
the first condition of monotonicity implies that $q^{k}_{s}(r_{1},...,r_{k-1})\geq q^{k}_{s}(\tilde{r}_{1},...,\tilde{r}_{k-1})$. By the size monotonicity, we have\[
		\mid Z_{k}\mid=\mid C_{s}^{k}(Z^{k-1},q_{s}^{k}(\tilde{r}_{1},...,\tilde{r}_{k-1}))\mid\geq\mid C_{s}^{k}(Y^{k-1},q_{s}^{k}(\tilde{r}_{1},...,\tilde{r}_{k-1}))\mid.
		\]
		
Moreover, by quota monotonicity, we obtain
\[
		\mid Y_{k}\mid-\mid C_{s}^{k}(Y^{k-1},q^{k}_{s}(\tilde{r}_{1},...,\tilde{r}_{k-1}))\mid\leq q^{k}_{s}(r_{1},...,r_{k-1})-q^{k}_{s}(\tilde{r}_{1},...,\tilde{r}_{k-1}).
\]
		
The difference on the left-hand side, $\mid Y_{k}\mid-\mid C_{s}^{k}(Y^{k-1},q^{k}_{s}(\tilde{r}_{1},...,\tilde{r}_{k-1}))\mid$, is the difference between the number of contracts chosen when the capacity is (weakly) increased from $q^{k}_{s}(\tilde{r}_{1},...,\tilde{r}_{k-1})$ to $q^{k}_{s}(r_{1},...,r_{k-1})$. Hence, it cannot exceed the increase in capacity, which is $q^{k}_{s}(r_{1},...,r_{k-1})-q^{k}_{s}(\tilde{r}_{1},...,\tilde{r}_{k-1})$. Therefore, we now have 
\[
		\mid Y_{k}\mid-\mid Z_{k}\mid\leq q^{k}_{s}(r_{1},...,r_{k-1})-q^{k}_{s}(\tilde{r}_{1},...,\tilde{r}_{k-1}).
\]
		
Rearranging gives us 
		\[
		q^{k}_{s}(\tilde{r}_{1},...,\tilde{r}_{k-1})-\mid Z_{k}\mid\leq q^{k}_{s}(r_{1},...,r_{k-1})-\mid Y_{k}\mid.
		\]
        
That is, $$\tilde{r}_{k}\leq r_{k}.$$
		
We now prove (2). Consider a contract $x\in Y^{k}$. That is, $x\notin C_{s}^{k}(Y^{k-1},q^{k}_{s}(r_{1},...,r_{k-1}))$. When the set $Y^{k-1}$ is expanded to $\tilde{Y}^{k-1}$, $x$ cannot be chosen since the choice function satisfies the substitutes property. That is,
\[
		x\notin C_{s}^{k}(Z^{k-1},q^{k}_{s}(r_{1},...,r_{k-1})).
\]

When the capacity is reduced to $q^{k}_{s}(\tilde{r}_{1},...,\tilde{r}_{k-1})$, the quota monotonicity property implies that 
\[
		x\notin C_{s}^{k}(Z^{k-1},q^{k}_{s}(\widetilde{r}_{1},...,\widetilde{r}_{k-1})).
\]
		
That is, $x$ cannot be chosen in category $k$. Hence, $x\in Z^{k}$. Thus, we conclude that $Y^{k}\subseteq Z^{k}$. 
		
Now, let $\tau_{j}=r_{j}-\tilde{r}_{j}$. As shown above, $\tau_{j}\geq0$
for all $j=1,...,K_{s}$. Putting in $$r_{j}=q_{s}^{j}(r_{1},...,r_{j-1})-\mid Y_{j}\mid$$ 
and 
$$\tilde{r}_{j}=q_{s}^{j}(\tilde{r}_{1},...,\tilde{r}_{k-1})-\mid Z_{j}\mid$$
we obtain the following.
\[
		\mid Z_{j}\mid=q_{s}^{j}(r_{1},...,r_{j-1})-q_{s}^{j}(\tilde{r}_{1},...,\tilde{r}_{j-1})+\mid Y_{j}\mid+\tau_{j}.
\]
        
Summing both the right- and left-hand sides for $j=1,...,K_{s}$ yields
\[
		\sum_{j=1}^{K_{s}} {\mid Z_{j}\mid}=\sum_{j=1}^{K_{s}}{\mid Y_{j}\mid}+\sum_{j=2}^{K_{s}}[q_{s}^{j}(r_{1},...,r_{j-1})-q_{s}^{j}(\tilde{r}_{1},...,\tilde{r}_{j-1})]+\sum_{j=1}^{K_{s}}{\tau_{j}}.
\]
		
Since each $\tau_{j}\geq0$, we have 
\[
		\sum_{j=1}^{K_{s}}{\mid Z_{j}\mid} \geq \sum_{j=1}^{K_{s}}{\mid Y_{j}\mid}+\sum_{j=2}^{K_{s}}{[q_{s}^{j}(r_{1},...,r_{j-1})-q_{s}^{j}(\tilde{r}_{1},...,\tilde{r}_{j-1})]}.
\]
		
Since $q_{s}^{j}(r_{1},...,r_{j-1})\geq q_{s}^{j}(\tilde{r}_{1},...,\tilde{r}_{j-1})$
for all $j=2,...,K_{s}$, we obtain the following. 
\[
		\sum_{[j=1}^{K_{s}}{\mid Z_{j}\mid} \geq \sum_{j=1}^{K_{s}}{\mid Y_{j}\mid}.
\]
        
That is, $$\mid\overline{C}_{s}(Y;q_{s})\mid\leq\mid\overline{C}_{s}(Z;q_{s})\mid.$$

\subsection*{Proof of Proposition 1}

We need to show that $C_{s}^{mT}$ is in the GSq family. The transfer functions under $C_{s}^{mT}$ satisfy monotonicity because $q_{s}^{D}=r_{OBC}$. \cite{fleiner2001matroid} shows that the first Step 1 of $C^{m}$ satisfies the substitute property. Moreover, \cite{sonmez2022affirmative} prove that $C^{m}$ satisfies size monotonicity. We then only need to show that $C^{m}$ satisfies quota-monotonicity. That is, for any given set of applicants $A\subseteq\mathcal{I}$ and $q\in\mathbb{N}$, we need to show that 

\begin{enumerate}
    \item $C^{m}\left(A,q\right)\subseteq C^{m}\left(A,q+1\right)$, and 
    \item $\mid C^{m}\left(A,q+1\right)\mid-\mid C^{m}\left(A,q\right)\mid\leq1$.
\end{enumerate}

We consider two cases: 	
	
\textbf{Case 1: The additional position is not reserved horizontally.} 

In this case, since the set of individuals $A\subseteq\mathcal{I}$ and the set of horizontally reserved positions $H^{v}=\bigcup_{j=1}^{L }H_{j}^{v}$ remain unchanged, when the capacity is increased, the horizontal type matching remains the same in Step 1 of $C^{m}$. Moreover, since in Step 2 the remaining positions are filled following the merit scores ranking with individuals among the unassigned individuals until all positions are assigned or all individuals are selected, every
individual who was chosen when capacity was $q$ will be selected when capacity increases to $q+1$. Therefore, both conditions of quota monotonicity are trivially satisfied.

\textbf{Case 2: The additional position is reserved horizontally.} 

Let $h_{j}$ be the horizontal type whose capacity increases by 1, w.l.o.g. Consider Step 1 (the Greedy process) under $C^{m}(A,q)$ and $C^{m}(A,q+1)$. Let $n^*$ be the first step in which the two choice procedures diverge. That is, the $n^*$-highest scoring individual is not selected under $C^{m}(A,q)$, but is chosen under $C^{m}(A,q+1)$. Note that in Steps $1.1$ to $1.(n^{*}-1)$, the same set of students are chosen in $C^{m}(A,q)$ and $C^{m}(A,q+1)$. If $i_{n^*}\notin C^{m}(A,q)$ and $i_{n^*}\in C^{m}(A,q+1)$, then it must be the case that $i_{n^*}$ is matched to a position reserved for the horizontal type $h_j$. 

Note also that the set of remaining students is and the set of horizontally reserved positions are the same at the end of Step $1.n^*$ of the greedy algorithm. Thus, the in the remaining steps, the same set of students will be chosen. Thus, each student who is chosen under $C^{m}(A,q)$ will continues to be chosen under $C^{m}(A,q+1)$. That is, $$C^{m}\left(A,q\right)\subseteq C^{m}\left(A,q+1\right).$$

Since each applicant chosen when capacity was $q$ continues to be chosen when capacity increases by one, the number of applicants chosen in Step 1 of $C^{m}$ stays the same or increases by one. By Step 2 of $C^{m}$, we can conclude that $$\mid C^{m}\left(A,q+1\right)\mid-\mid C^{m}\left(A,q\right)\mid\leq1.$$

\subsection*{Proof of Proposition 2}
The choice rules used in Step 1 and Step 2 are both q-responsive induced by applicants test scores. Thus, they satisfy the substitutes property, size monotonicity, and quota monotonicity. The capacity of the open category positions that are allocated in Step 2 is $q_{s}^{o}+ \sum_{m\in M}r^{m}_{s}$ where $r_{s}^{m}$ denotes the number of unfilled seats that were reserved for middle school $m\in M$. This transfer policy straightforwardly satisfy monotonicity condition. Thus, $\mathcal{C}^{SimRO}$ is a GSq choice rule according to Definition 3. 

\subsection*{Proof of Proposition 3}
The choice rules used in Steps 1-3 are all q-responsive induced by test scores. Therefore, they satisfy the substitutes property, size monotonicity, and quota monotonicity. The capacity of the open category positions that are allocated in Step 3 is $\sum_{m\in M}r^{m}_{s}$, where $r_{s}^{m}$ denotes the number of unfilled seats that were reserved for middle school $m\in M$. This transfer policy is monotonic. Then, by Definition 3, $\mathcal{C}^{SimORO}$ is a GSq choice rule.

\subsection*{Proof of Proposition 4}

Given $I,J\subset \mathcal{I}$ such that $I \cap J=\emptyset$, $\mathcal{C}^{SimSep}(I,J)$ assigns the top $q_s^o$ applicants within the set $I$ to open positions and the top $q_s^m$ applicants within $J \cap \mathcal{I}^{m}$ to reserved positions for $m\in M$.

$\mathcal{C}^{SimSep}$ satisfies the subsitutes condition and size monotonicity. Thus, it belongs to the GSq family.

\subsection*{Proof of Proposition 5}

Given $I,J\subset \mathcal{I}$ such that $I \cap J=\emptyset$, $\mathcal{C}^{SimFlex}(I,J)$ assigns the top $q_s^m$ applicants within the set $J$ to positions reserved for $m\in M$ in the first step. In the second step, it considers all remaining positions and assigns the top $q_s^o$ applicants within $I$ to open positions.

$\mathcal{C}^{SimSep}$ satisfies the subsitutes condition and is q-acceptant. Hence, it satisfies size monotonicity. Therefore, it belongs to the GSq family.

\subsection*{Proof of Proposition 6}
\cite{hu2025verifiable} show that $\mathcal{C}^{SimOR}$ satisfies the subsititues condition and is $q$-acceptant. Since $q$-acceptance implies size monotonicity, it also satisfies size monotonicity. Since $\mathcal{C}^{SimOR}$ satisfies the substitutes property and size monotonicity, it can be regarded as a single division's choice rule. Thus, it is a GSq choice rule.

\newpage

\bibliographystyle{econometrica}
\bibliography{bertan}

@article{aygun2025affirmative,
	author = {Ayg{\"u}n, Orhan and Turhan, Bertan},
	journal = {Iowa State University Working Paper},
	title = {Affirmative Action in India: A Market Design Approach with Historical and Legal Perspective},
	year = {2025}}

@article{romm2024stability,
  title={Stability vs. no justified envy},
  author={Romm, Assaf and Roth, Alvin E and Shorrer, Ran I},
  journal={Games and Economic Behavior},
  volume={148},
  pages={357--366},
  year={2024},
  publisher={Elsevier}
}

@article{gul1999walrasian,
  title={Walrasian equilibrium with gross substitutes},
  author={Gul, Faruk and Stacchetti, Ennio},
  journal={Journal of Economic Theory},
  volume={87},
  number={1},
  pages={95--124},
  year={1999},
  publisher={Elsevier}
}

@article{evren2025reserve,
  title={Reserve Matching with Thresholds},
  author={Evren, Suat},
  journal={arXiv preprint arXiv:2309.13766},
  year={2025}
}

@article{hu2025verifiable,
  title={Verifiable affirmative action in Chinese high school admissions},
  author={Hu, Xinquan and Zhang, Jun},
  journal={arXiv preprint arXiv:2504.04689},
  year={2025}
}

@article{aygun2013matching,
	title={Matching with contracts: Comment},
	author={Ayg{\"u}n, Orhan and S{\"o}nmez, Tayfun},
	journal={American Economic Review},
	volume={103},
	number={5},
	pages={2050--2051},
	year={2013},
	publisher={American Economic Association}
}

@article{echenique2007counting,
  title={Counting combinatorial choice rules},
  author={Echenique, Federico},
  journal={Games and Economic Behavior},
  volume={58},
  number={2},
  pages={231--245},
  year={2007},
  publisher={Elsevier}
}

@article{roth1999redesign,
  title={The redesign of the matching market for American physicians: Some engineering aspects of economic design},
  author={Roth, Alvin E and Peranson, Elliott},
  journal={American Economic Review},
  volume={89},
  number={4},
  pages={748--780},
  year={1999},
  publisher={American Economic Association}
}

@article{greenberg2024redesigning,
  title={Redesigning the US Army’s Branching Process: A Case Study in Minimalist Market Design},
  author={Greenberg, Kyle and Pathak, Parag A and S{\"o}nmez, Tayfun},
  journal={American Economic Review},
  volume={114},
  number={4},
  pages={1070--1106},
  year={2024},
  publisher={American Economic Association 2014 Broadway, Suite 305, Nashville, TN 37203}
}

@article{aziz2024efficient,
  title={Efficient and Fair Healthcare Rationing},
  author={Aziz, Haris and Brandl, Florian},
  journal={Journal of Artificial Intelligence Research},
  volume={81},
  pages={337--358},
  year={2024}
}

@article{roth1990new,
  title={New physicians: a natural experiment in market organization},
  author={Roth, Alvin E},
  journal={Science},
  volume={250},
  number={4987},
  pages={1524--1528},
  year={1990},
  publisher={American Association for the Advancement of Science}
}

@article{pathak2024fair,
  title={Fair Allocation of Vaccines, Ventilators and Antiviral Treatments: Leaving No Ethical Value Behind in Healthcare Rationing},
  author={Pathak, Parag and Sönmez, Tayfun and Ünver, Utku and Yenmez, Bumin},
  journal={Management Science},
  volume={70},
  issue={6},
  pages={3381-4165},
  year={2024}
}

@article{sonmez2022affirmative,
	title={Affirmative action in India via vertical, horizontal, and overlapping reservations},
	author={S{\"o}nmez, Tayfun and Yenmez, M Bumin},
	journal={Econometrica},
	volume={90},
	number={3},
	pages={1143--1176},
	year={2022},
	publisher={Wiley Online Library}
}

@article{biro2020need,
  title={Need versus merit: the large core of college admissions markets},
  author={Bir{\'o}, P{\'e}ter and Hassidim, Avinatan and Romm, Assaf and Shorrer, Ran I and S{\'o}v{\'a}g{\'o}, S{\'a}ndor},
  journal={arXiv preprint arXiv:2010.08631},
  year={2020}
}

@article{hassidim2021limits,
  title={The limits of incentives in economic matching procedures},
  author={Hassidim, Avinatan and Romm, Assaf and Shorrer, Ran I},
  journal={Management Science},
  volume={67},
  number={2},
  pages={951--963},
  year={2021},
  publisher={INFORMS}
}

@inproceedings{fleiner2001matroid,
	title={A matroid generalization of the stable matching polytope},
	author={Fleiner, Tam{\'a}s},
	booktitle={International Conference on Integer Programming and Combinatorial Optimization},
	pages={105--114},
	year={2001},
	organization={Springer}
}

@article{hatfield2019hidden,
	title={Hidden substitutes},
	author={Hatfield, John William and Kominers, Scott Duke},
	journal={Working Paper},
	year={2019}
}

@article{aygun2022dereserve,
	author = {Ayg{\"u}n, Orhan and Turhan, Bertan},
	journal = {Management Science},
	volume = {69},
    number = {10},
	pages = {5695--6415},
	title = {How to de-reserve reserves: Admissions to technical colleges in India},
	year = {2022}}

@article{aygun2021college,
	title={College admission with multidimensional privileges: The Brazilian affirmative action case},
	author={Ayg{\"u}n, Orhan and B{\'o}, In{\'a}cio},
	journal={American Economic Journal: Microeconomics},
	volume={13},
	number={3},
	pages={1--28},
	year={2021},
	publisher={American Economic Association 2014 Broadway, Suite 305, Nashville, TN 37203-2425}
}

@article{kojima2012school,
		title={School choice: Impossibilities for affirmative action},
		author={Kojima, Fuhito},
		journal={Games and Economic Behavior},
		volume={75},
		number={2},
		pages={685--693},
		year={2012},
		publisher={Elsevier}
	}

@article{hatfield2017stable,
	title={Stable and strategy-proof matching with flexible allotments},
	author={Hatfield, John William and Kominers, Scott Duke and Westkamp, Alexander},
	journal={American Economic Review},
	volume={107},
	number={5},
	pages={214--219},
	year={2017}
}

@article{hatfield2017restud,
	title={Stability, strategy-proofness, and cumulative offer mechanisms},
	author={Hatfield, John William and Kominers, Scott Duke and Westkamp, Alexander},
	journal={Review of Economic Studies},
	volume={88},
	number={3},
	pages={1457--1502},
	year={2021}
}

@article{gale1962college,
	Author = {Gale, David and Shapley, Lloyd S},
	Journal = {The American Mathematical Monthly},
	Number = {1},
	Pages = {9--15},
	Publisher = {JSTOR},
	Title = {College admissions and the stability of marriage},
	Volume = {69},
	Year = {1962}}

@article{hafalir2013effective,
	Author = {Hafalir, Isa E and Yenmez, M Bumin and Yildirim, Muhammed A},
	Journal = {Theoretical Economics},
	Number = {2},
	Pages = {325--363},
	Publisher = {Wiley Online Library},
	Title = {Effective affirmative action in school choice},
	Volume = {8},
	Year = {2013}
}

@article{kelso1982job,
	title={Job matching, coalition formation, and gross substitutes},
	author={Kelso, Alexander S and Crawford, Vincent P},
	journal={Econometrica},
	pages={1483--1504},
	year={1982},
	publisher={JSTOR}
}

@article{chambers2017choice,
	title={Choice and matching},
	author={Chambers, Christopher P and Yenmez, M Bumin},
	journal={American Economic Journal: Microeconomics},
	volume={9},
	number={3},
	pages={126--147},
	year={2017},
	publisher={American Economic Association 2014 Broadway, Suite 305, Nashville, TN 37203-2425}
}

@article{hatfield2005matching,
	title={Matching with contracts},
	author={Hatfield, John William and Milgrom, Paul R},
	journal={American Economic Review},
	volume={95},
	number={4},
	pages={913--935},
	year={2005},
	publisher={American Economic Association}
}

@article{alkan2003stable,
	title={Stable schedule matching under revealed preference},
	author={Alkan, Ahmet and Gale, David},
	journal={Journal of Economic Theory},
	volume={112},
	number={2},
	pages={289--306},
	year={2003},
	publisher={Elsevier}
}

@article{kominers2016matching,
	title={Matching with slot-specific priorities: Theory},
	author={Kominers, Scott Duke and S{\"o}nmez, Tayfun},
	journal={Theoretical Economics},
	volume={11},
	number={2},
	pages={683--710},
	year={2016},
	publisher={Wiley Online Library}
}

@article{aygun2023priority,
	title={Priority design for engineering colleges in India},
	author={Ayg{\"u}n, Orhan and Turhan, Bertan},
	journal={Indian Economic Review},
    volume = {58},
	pages={5--15},
	year={2023}
}

@article{baswanaetal2019,
	title={Centralized admissions for engineering colleges in India},
	author={Baswana, Surender and Chakrabarti, Partha Pratim and Chandran, Sharat and Kanoria, Yashodhan and Patange, Utkarsh},
	journal={INFORMS Journal on Applied Analytics},
	volume={45},
	number={5},
	pages={338--354},
	year={2019},
}

@article{alva2023choice,
	title={Choice and Market Design},
	author={Alva, Samson and Dogan, Battal},
	journal={Online and Matching-Based Market Design},
	pages={238-263},
	year={2023},
	publisher={Cambridge University Press}
}

@article{ehlers2014school,
	title={School choice with controlled choice constraints: Hard bounds versus soft bounds},
	author={Ehlers, Lars and Hafalir, Isa E and Yenmez, M Bumin and Yildirim, Muhammed A},
	journal={Journal of Economic Theory},
	volume={153},
	pages={648--683},
	year={2014},
	publisher={Elsevier}
}

@article{aygun2017large,
	title={Large-scale affirmative action in school choice: Admissions to IITs in India},
	author={Ayg{\"u}n, Orhan and Turhan, Bertan},
	journal={American Economic Review},
	volume={107},
	number={5},
	pages={210--213},
	year={2017},
	publisher={American Economic Association 2014 Broadway, Suite 305, Nashville, TN 37203}
}

@article{aygun2020dynamic,
	title={Dynamic reserves in matching markets},
	author={Ayg{\"u}n, Orhan and Turhan, Bertan},
	journal={Journal of Economic Theory},
	volume={188},
	pages={105069},
	year={2020}
}

@article{kamada2015efficient,
	title={Efficient matching under distributional constraints: Theory and applications},
	author={Yuichiro Kamada and Fuhito Kojima},
	journal={American Economic Review},
	volume={105},
	number={1},
	pages={67--99},
	year={2015}
}

@article{westkamp2013analysis,
	Author = {Westkamp, Alexander},
	Journal = {Economic Theory},
	Number = {3},
	Pages = {561--589},
	Publisher = {Springer},
	Title = {An analysis of the German university admissions system},
	Volume = {53},
	Year = {2013}}

@article{echenique2015control,
	author = {Echenique, Federico and Yenmez, M Bumin},
	journal = {American Economic Review},
	number = {8},
	pages = {2679--94},
	title = {How to control controlled school choice},
	volume = {105},
	year = {2015}}

\end{document}